\documentclass[titlepage,oneside,letterpage,11pt]{article}

\usepackage[tiny,rm]{titlesec}
\newpagestyle{trbstyle}{
	\sethead{Le and Ukkusuri}{}{\thepage}
}

\titleformat{\section}{\bfseries}{}{0pt}{\uppercase}
\titlespacing*{\section}{0pt}{11pt}{*0}
\titleformat{\subsection}{\bfseries}{}{0pt}{}
\titlespacing*{\subsection}{0pt}{11pt}{*0}
\titleformat{\subsubsection}{\itshape}{}{0pt}{}
\titlespacing*{\subsubsection}{0pt}{11pt}{*0}

\usepackage{mathtools}
\usepackage[ruled,linesnumbered]{algorithm2e}

\usepackage{graphicx}
\usepackage{epstopdf}
\usepackage{multirow}
\usepackage{paralist}
\usepackage{hyperref}
\usepackage{amsmath,amsthm,amssymb}
\usepackage{setspace}
\usepackage{float}
\usepackage{morefloats}

\usepackage[figuresright]{rotating}

\usepackage{color}
\usepackage{longtable}
\usepackage{multirow}
\usepackage{cases}

\usepackage[]{algorithm2e}
\usepackage{mdwlist}
\usepackage{epstopdf}
\usepackage{booktabs} 
\usepackage{amsfonts}
\usepackage{paralist}
\newcommand{\be}{\begin{equation}}
\newcommand{\ee}{\end{equation}}

\newcommand{\upperRomannumeral}[1]{\uppercase\expandafter{\romannumeral#1}}

\makeatletter
\newcommand{\removelatexerror}{\let\@latex@error\@gobble}
\makeatother

\usepackage{enumitem}
\setlist[1]{labelindent=0.5in,leftmargin=*}
\setlist[2]{labelindent=0in,leftmargin=*}
\setlist{nosep} 

\setlist[itemize]{label=\textbullet}

\usepackage{ccaption}
\usepackage{amsmath}
\makeatletter
\renewcommand{\fnum@figure}{\textbf{FIGURE~\thefigure} }
\renewcommand{\fnum@table}{\textbf{TABLE~\thetable} }
\makeatother
\captiontitlefont{\bfseries \boldmath}
\captiondelim{\;}

\usepackage{mathptmx}

\usepackage[sort&compress,numbers]{natbib}
\renewcommand{\cite}[1]{({\it \citenum{#1}})}

\setcitestyle{round}

\usepackage[pagewise]{lineno}

\newread\somefile
\usepackage{xparse}

\newcounter{totalwordcounter}
\newcounter{wordcounter}
\makeatletter

\NewDocumentCommand{\wordcount}{s}{%
	\immediate\write18{texcount -sum -1 \jobname.tex > count.txt}%
	\immediate\openin\somefile=count.txt%
	\read\somefile to \@@localdummy%
	\immediate\closein\somefile%
	\setcounter{wordcounter}{\@@localdummy}%
	\IfBooleanF{#1}{%
		\@@localdummy
	}%
}
\makeatother

\usepackage{totcount}
\regtotcounter{table} 	
\regtotcounter{figure} 	

\newcommand{\wordfigure}{250} 
\newcommand{\wordtable}{250} 

\newcommand{\totalwordcount}{%
	\wordcount*
	\setcounter{totalwordcounter}{\value{wordcounter}}%
	\addtocounter{totalwordcounter}{\numexpr\wordfigure*\totvalue{figure}}%
	\addtocounter{totalwordcounter}{\numexpr\wordtable*\totvalue{table}} %
	\number\value{totalwordcounter}
	\renewcommand{\totalwordcount}{\number\value{totalwordcounter}}
}

\usepackage[export]{adjustbox}

\usepackage{graphicx}
\usepackage{pbox}
\usepackage{latexsym}
\usepackage[flushleft]{threeparttable}

\usepackage{color,soul}
\usepackage{subcaption}
\usepackage{rotating}
\usepackage{array}
\usepackage{tabulary}
\usepackage[pagewise]{lineno}
\usepackage{pdflscape}
\usepackage{makecell,   
	multirow, longtable}
\setcellgapes{3pt}

\usepackage{siunitx}    

\usepackage{amssymb}
\usepackage{amsthm}

\def\<#1>{\textit{#1}}

\usepackage{comment}

\usepackage{graphicx}
\usepackage{pbox}
\usepackage{latexsym}

\usepackage{color,soul}
\usepackage{subcaption}
\usepackage{rotate}
\usepackage{rotating}
\usepackage{array}
\usepackage{tabulary}
\usepackage[pagewise]{lineno}

\usepackage{makecell,   
	multirow, longtable}
\setcellgapes{3pt}
\usepackage{array,multirow,graphicx}

\usepackage{threeparttable}
\usepackage{booktabs}
\usepackage{threeparttablex}
\usepackage{geometry}
\usepackage{pdflscape}

\usepackage{siunitx}    

\usepackage[hyphenbreaks]{breakurl}
\usepackage{amssymb}
\usepackage{amsthm}

\def\<#1>{\textit{#1}}

\usepackage{comment}
\usepackage{paralist}

\usepackage{caption}
\usepackage[justification=centering]{caption}

\usepackage{mathtools}
\usepackage{comment}

\input{glyphtounicode}
\usepackage[utf8]{inputenx}
\usepackage[T1]{fontenc}

\usepackage{booktabs}

\begin{document}
	\thispagestyle{empty}
	
	
	\begin{titlepage}
		\begin{flushleft}
			
			{\bfseries{INFLUENCING FACTORS THAT DETERMINE THE USAGE OF THE CROWD-SHIPPING SERVICES}}\\[36pt]
								
			{\bfseries Tho V. Le} \\
			Ph.D. Candidate \\
			Lyles School of Civil Engineering\\
			Purdue Univeristy\\
			550 Stadium Mall Drive\\
			West Lafayette, IN 47907\\
			le39@purdue.edu\\
             [11pt]
                        		
			{\bfseries Satish V. Ukkusuri, Ph.D. }{Corresponding author}\\
			Professor\\
			Lyles School of Civil Engineering\\
			Purdue Univeristy\\
			550 Stadium Mall Drive\\
			West Lafayette, IN 47907\\
			sukkusur@purdue.edu\\[11pt]
					

			Submission Date: February 15$^{th}$, 2019
		\end{flushleft}
	\end{titlepage}
	
	
	\newpage

\section{Abstract}
\noindent The objective of this study is to understand how senders choose shipping services for different products, given the availability of both emerging crowd-shipping (\<CS>) and traditional carriers in a logistics market. Using data collected from a US survey, Random Utility Maximization (RUM) and Random Regret Minimization (RRM) models have been employed to reveal factors that influence the diversity of decisions made by senders. Shipping costs, along with additional real-time services such as courier reputations, tracking info, e-notifications, and customized delivery time and location, have been found to have remarkable impacts on senders' choices. Interestingly, potential senders were willing to pay more to ship grocery items such as food, beverages and medicines by \<CS> services. Moreover, the real-time services have low elasticities, meaning that only a slight change in those services will lead to a change in sender-behavior. Finally, data-science techniques were used to assess the performance of the RUM and RRM models and found to have similar accuracies. The findings from this research will help logistics firms address potential market segments, prepare service configurations to fulfill senders' expectations, and develop effective business operations strategies. \\

\noindent \<Keywords>: Crowd-shipping, Random regret, Random utility, Willingness to pay, Data science.
	

\newpage



\section{Introduction}

\noindent Across many industries, numerous sharing-economy companies have been established to provide common platforms for directly connecting supply and demand. However, only a fraction of the sharing-economy startups succeed at expanding their market share and maintaining customers \citep{Dablanc2016}. One outstanding example of a sharing-economy startup is AirBnB \citep{li2015agent,ke2017service,zervas2017rise}. AirBnB's value increased from \$10 billion in 2014 to \$31 billion in 2017 \citep{Statista2018}. On the other hand, Uber, a popular car-sharing platform, offers various services supporting either passenger mobility (e.g., economy, premium, accessibility, and carpool), food delivery (e.g., Uber Eats), or freight transport (e.g., Uber Freight) (https://www.uber.com/). However, out of those services, only UberEats generates profits \citep{Isaac2017}. To recognize the supply-demand interactions, an important question is to understand the underlying factors that contribute to the growth of these services.

With the development of new Internet- and mobile-connected technologies, the last-mile delivery market has transformed into a shared market with both crowd-shipping (\<CS>) and traditional logistics carriers (\<TLCs>) competing with each other for shipments. \<CS> is an example of the sharing economy in the logistics industry. The main idea of \<CS> services is to encourage crowds to transport goods whenever they have flexibility or an opportunity. The "crowd" may be individuals or agencies. The goods can be transported by personal cars, bikes, buses, metros, taxis, or even pedestrians. Therefore, \<CS> may differ from \<TLCs> in operations and fare setting, bringing competition to the traditional logistics industry. 

In a competitive market, senders have more delivery options. Therefore, they show varied behaviors in different shipping contexts. Senders, who are not a homogeneous group, have various tastes for services. Some senders may be willing to pay more to get their packages to be delivered faster. Some senders may be willing to pay more for a certain shipment but not for others. Another subset of senders may never opt for personalized services, only wanting the lowest possible cost. Furthermore, for different shipment classes, such as urgent purchases, tight-window delivery packages, peripheral products, or durable goods, senders may have to trade-off between service options and show various shipping behaviors. Some senders may want to get the maximum utility for the money they spend. Other senders, however, may try to minimize the regret of selecting not the best courier option (e.g. service) for sending their shipment. 
As such, the following questions have been raised: What factors influence senders' choices of couriers for different types of shipments? Which choice theory, maximum utility or minimum regret, do senders rely on to select a courier? How much are senders willing to pay for the last-mile delivery for different types of shipments? What are the relationships between an individual's socio-demographic characteristics, shipment types, and willingness to pay (\<WTP>)? Which products are more likely to be sent via \<CS> services?

The current situation suggests flexible and cost-effective delivery systems that utilize existing infrastructure and delivery services are needed, especially for the first- and last-mile delivery. Whenever a new service plans to be released or has just launched in the market, decision-making behaviors should be investigated to understand customers' expectations and satisfaction-levels in order to improve the service and make it more competitive in the marketplace. This research presents a set of factors influencing senders' choices of delivery couriers in the context of a shared market with the coexistence of \<CS> and \<TLCs>. In addition, the senders' \<WTP> for certain shipping products and service features will also be studied. Insights from this study provide a better understanding of demand behaviors under various service levels which helps \<CS> companies fine-tune their systems to satisfy senders' expectations.

This paper is organized into seven sections. Section one introduces the context and motivations for the research. Section two reviews studies relating to the research topic. Section three explains our research approaches including questionnaire designs, data collections, and data descriptions. Section three also shows the modeling approaches including mixed logit and regret minimization models as well as our \<WTP> computational method. Section four illustrates estimation results and insights. Section five discusses simulation results including senders' perceptions and elasticity analysis. Section six presents data-science testing techniques for evaluating our models' accuracy. The paper is concluded in section seven.

\section{Related studies}

\noindent The sharing economy starts with the idea of renting out useful assets or services for a short period of time. Though it has become popular in recent years, the sharing economy's overall benefits are controversial and not yet clear \citep{dervojeda2013sharing,schor2016debating,frenken2017putting}. Since there are limited studies of the demand side of a logistics market in which both \<CS> and \<TLCs> are available, in this section we will include in our review several studies that deal separately with the sharing economy and traditional logistics companies. This review does not aim to be exhaustive but rather  to provide a big picture. With our review, we can still generate general ideas about factors influencing demand and shaping consumer behavior--ideas which will be valuable and applicable to \<CS> firms.

\subsection{Traditional logistics}
\noindent Mode and service choices in traditional logistics have been well-studied. Factors commonly found to influence consumer behavior include price, delivery time, reliability \citep{train2008estimation,anderson2009demand}, flexibility, frequency \citep{danielis2007attribute,zamparini2011monetary}, carrier reputation, courier reputation, customer services, billing accuracy, facility/equipment availability, and intact packages \citep{cavalcante2013shipper}.

With the emergence of e-commerce, however, traditional logistics companies have been challenged by decentralized orders and new requirements for flexibility and on-demand shipping. In the report of \citet{Joerss2016}, the three product categories which saw the most e-shoppers declining their purchases due to long delivery time were "groceries" (27\%), "medications" (26\%), and "books, CDs, DVDs, and video games" (20\%). It is apparent that delivery services can significantly affect e-shoppers' purchasing decisions.

\subsection{Peer-to-peer accommodations and AirBnB}
\noindent \citet{tussyadiah2016factors} collected data from approximately 650 users of peer-to-peer accommodations to investigate factors influencing users' satisfaction-levels. It turned out that the most salient factors were happiness, cost-savings, and household amenities. Furthermore, the "social benefits'' factor was influential for private-room renters but not whole-home renters.

In recent years, AirBnB has sharply extended its market, recording 140 million guests using AirBnB services from 2008 to 2016, 120 million of those guests using the service in just 2015 and 2016 \citep{AirBnBcitizen2017}. \citet{guttentag2017tourists} found that a majority of travelers using AirBnB services were motivated by ``interaction,'' ``home benefits,'' ``novelty,'' ``sharing economy ethos,'' and ``local authenticity.''

\subsection{Ride-sharing}
\noindent For ride-sharing services, a few studies have outlined users' preferences compared to traditional taxi services and public transportation. \citet{oliphant2008native} found that saving gasoline and time, environmental friendliness, and flexibility were the main motivations for the Washington, DC ride-share users. Moreover, ride-sharing users in San Francisco were motivated by payment convenience, shorter waiting times, and faster service as reported by \citet{rayle2016just}. Similarly, \citet{shaheen2016casual} found convenience (e.g., pickup point), cost and time saving as well as age and job situation drove casual carpooling or other mode-choice behaviors.  
Not to mention, \citet{dias2017behavioral} learned that users were more likely influenced to use or not to use car-sharing or ride-sourcing services by their age, education level, income, employment status, and residential density. In fact, \citet{Smith2016} has found urban residents were major users of the US e-hailing services. Likewise, \citet{carrese2017real} revealed gender, convenience (e.g., personalization on pickup time and location), and comfort interacting with others significantly influenced ride-sharing system-users in the Lazio region of Italy.

In a study by \citet{Bennett2015}, about two-thirds of Uber and Lyft users were in the age group of 25-55 years old. In that same study, women accounted for  52\% and 58\% of Uber and Lyft users, respectively. Users, across gender, liked that the services were convenient (e.g., can be e-hailed from a smart phone), had shorter waiting times, could be tracked in real-time, were cost-competitive, and could be advised in advance. Scraping data from Yahoo emails, 
\citet{kooti2017analyzing} studied Uber users and found race, age, gender, income, ride-sharing service (e.g., UberX), matching strategy (e.g., to similar-aged driver), the time of the day (e.g., night time), the day of the week (e.g., weekend), and promotions remarkably affected riders' activities once using the services.
In a nutshell, socio-demographic, service, and contextual factors all significantly influence ride- and car-sharing users. Service levels, nevertheless, have only been investigated in a couple of studies.

\subsection{Crowd-shipping}
\noindent \<CS> systems provide potential benefits for a wide range of stakeholders. \<CS> users can benefit from shorter delivery times, cheaper delivery costs, and increased flexibility and accessibility. Retailers, on the other hand, may have more delivery options, spend less for delivery, and attract more customers due to using better delivery services. When \<CS> strategies are properly applied, societies can reduce traffic congestion, pollution, and accidents while increasing employment opportunities. Nonetheless, the young \<CS> industry is also facing several challenges, such as the chicken-or-the-egg problem. The chicken is the \<CS> services, while the egg is the demand for sending packages via the \<CS> system. To overcome this challenge and achieve potential benefits, it is necessary to understand stakeholder preferences, particularly the \<CS> users' behaviors. 

While researchers have made significant efforts to collect data and evaluate \<CS> systems, last-mile delivery research has been limited by data availability, especially that of behavioral data. One thing that data has revealed, however, is that a limited number of people currently know about or have already used \<CS> services \citep{miller2017crowdsourced,punel2017modeling}. Therefore, it is critical to figure out what needs to be done to attract more users to \<CS> systems. 

There are a few studies focusing on the topic of the \<CS> demand. \citet{briffaz2016crowd} found about three-fourths of respondents were potential \<CS> service-users. Top factors which influenced respondents' choices were delivery cost (32\%), flexibility (29\%), and environmental benefits (22\%). Additionally, \citet{Le2017} found potential \<CS> users were concerned about damages and delivery-time punctuality. Users preferred to have packages delivered during late afternoon or early evening on weekdays, or from mid-morning to late afternoon on weekends. 
\citet{ballare2018preliminary} studied data provided from a \<CS> company and found age as well as package size, shipping distance, and service price were main influenced factors on \<CS> users. In another research, \citet{punel2018} investigated data collected from an on-line survey and found young men and full-time employees were more likely to use \<CS> services. 
Elsewhere, \citet{punel2017modeling} investigated the effects of contexts (i.e. distance) and experience on users' acceptance of \<CS>. Public acceptance varied according to distance. For instance, short-distance users were more concerned about delivery speeds and couriers' reputations. Long-distance users, differently, preferred goods to be delivered by experienced couriers and via higher service levels. Not to mention, \citet{punel2017modeling} also computed \<WTP> and sensitivity for \<CS> services' attributes. 
In sum, all available studies on the \<CS> demand side only presented the results of descriptive analysis except for the study of \citet{punel2017modeling}.

In Table \ref{tab4.RelatedStudies4}, factors influencing users of AirBnB, ride-sharing, traditional logistics, and \<CS> are summarized. Those factors can be categorized into personal characteristics, service levels, post-purchase services, couriers' qualifications, contexts, social interactions, and social benefits. 

\begin{table}
	\centering
	\caption{Attributes for alternative choices from existing studies}
	\scriptsize
	\label{tab4.RelatedStudies4}               
	\begin{tabular}{p{1.5cm} p {4.5cm} p{8cm}}
		\hline
		\textbf{Area}	 		& \textbf{Study} 						& \textbf{Variables}\\ [0.5ex] 	
		\hline
		Traditional logistics 	& \citet{train2008estimation, anderson2009demand, danielis2007attribute, zamparini2011monetary, cavalcante2013shipper}		&   Price, delivery time, reliability, flexibility, frequency, carrier reputation, courier reputation, customer services, billing accuracy, facility/equipment availability, and intact packages. \\ \hline
		AirBnB					& \citet{tussyadiah2016factors,guttentag2017tourists} 			&  Less expenses, venue advantageous, various amenities, enjoyment, local authenticity, social interaction, and supporting for sharing economy.   \\ \hline  
		Ride- and car-sharing	& \citet{oliphant2008native,Bennett2015,rayle2016just,Smith2016,shaheen2016casual,carrese2017real,kooti2017analyzing,dias2017behavioral} 			& Race, age, gender, income, education level, employment status, residential density, time-of-day, day of week, cost saving and quoted in advance, travel time saving, waiting time saving, flexibility in pickup time and location, promotions, matching strategies, track live, payment convenience, interaction comfort, and environmental friendly.\\ \hline        
		Crowd-shipping			& \citet{briffaz2016crowd,Le2017,punel2017modeling}		&   Age, gender, income, education, employment, distance, delivery cost, delivery time, flexibility, intact products, reliability, environment benefits, and couriers' experience and reputation.  \\ \hline
		
	\end{tabular}
\end{table}

To the best of these authors' knowledge, there is no available study about the influence of shipment classes on courier choices in a logistics market featuring both \<CS> and \<TLCs> that uses random-parameter modeling. Moreover, safety and security concerns are claimed as challenges to \<CS> services. Senders may consider to minimize regret once shipping some products but to maximize utility once sending another product. Therefore, the contributions of this study include: 1) the development of random utility maximization and random regret minimization models for examining senders' choices; 2) Providing different sets of influencing factors on senders' choices of couriers for sending different shipment classes. The sets comprise of app-based attributes of logistics services and socio-demographic factors. Knowing the factors help \<CS> firms to prepare and provide tailored services for a given market segment; 3) Providing \<WTP> values which are helpful for making pricing decisions; 4) Supplying elasticities of decisions made by senders; and 5) Using data science techniques to validate RUM and RRM models' accuracies.

\section{Research approaches}

\begin{figure}
	\centering
	\includegraphics[width=0.5 \textwidth]{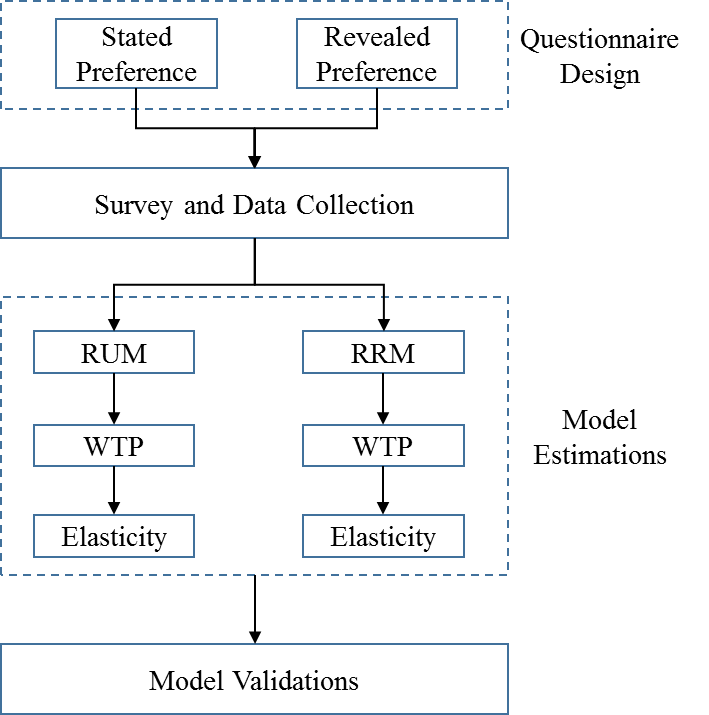}
	\caption{Research framework}
	\label{fig4.wtpframework}	
\end{figure}

This study follows a framework which is presented in Figure \ref{fig4.wtpframework}. As can be seen, there are four main steps, namely questionnaire design, survey and data collection, model estimations, and model validations. RUM and RRM stand for random utility maximization and random regret minimization, respectively. Each step will be discussed in great details in the next sub-sections.

\par

\subsection{Stated preference (SP) survey design and sample descriptions}

\subsubsection{Data source}
\noindent This study will first design a set of questionnaires featuring SP questions to examine senders' preferences when shipping packages. The attributes are designed to understand the senders' behaviors for selecting couriers. In the SP part, attributes of services (e.g., fare, delivery time, and shipment characteristics) are designed with multiple levels of services. The alternatives' attributes and corresponding meanings are summarized in Table \ref{tab4.Attribute_meaning}. Products commonly purchased from e-shopping, such as apparel, books, music, videos and consumer electronics, often require delivery from a courier service \citep{ATKearney2014a}. Those products, along with with fast foods, flowers, groceries, beverages, dried foods, personal health items, and medicines, are grouped into 8 categories and included in the final questionnaire. Couriers 1-3 have been designed to have shorter delivery time but higher delivery cost than those of courier 4.

\begin{table}
	\centering
	\caption{Alternatives' attributes and corresponding meanings}
	\small
	\label{tab4.Attribute_meaning}
	\begin{tabular}{p{6cm} p{8cm}} 
		\hline
		\textbf{Attributes}	& \textbf{Description} \\ [0.5ex] 
		\hline		
		Shipping cost          &              \$14, \$18, \$22, \$26 (4 levels) \\                
		Delivery time          &               1.5h, 3h, 5h, once/day, Delivery within 2-4 days (5 levels)     \\
		Courier's reputation/ ranking          &               High, Medium, Low (3 levels)                              \\
		Apps (tracking and tracing features)          &               Yes, No (2 levels)                       \\
		Apps (electronic delivery notification)          &               Yes, No (2 levels)                      \\     
		Personalization for delivery time window          &               Yes, No (2 levels)                       \\     
		Personalization for location of delivery          &               Home, Other (e.g., your car's trunk), Pickup at a carrier's location (3 levels)\\
		Payment method          &               On app/website (automatic), By cash (2 levels) \\
		Willingness to tip          &               No tip, \$1, \$2, \$3 (4 levels)             \\              
		\hline
	\end{tabular}
\end{table}

\begin{figure}
	\centering
	\includegraphics[width=1\textwidth]{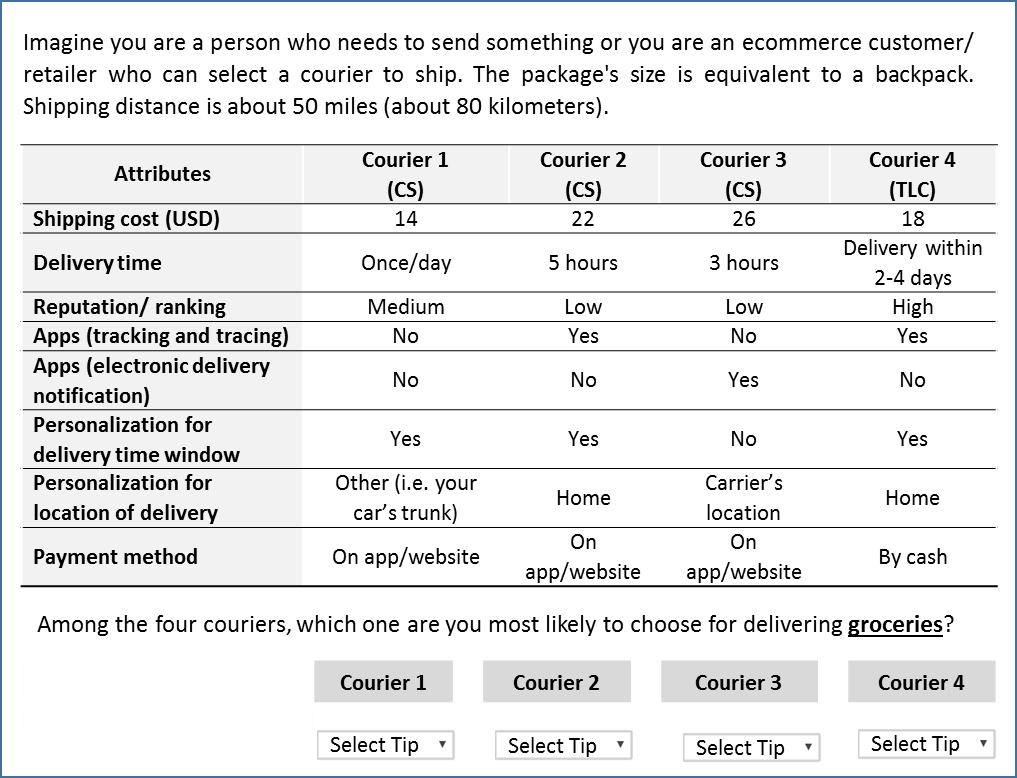}
	\caption{ Stated preference questions sample}
	\label{fig4.Question_sample}	
\end{figure}\par

The survey was created in Qualtrics which is an on-line survey platform. The link to the on-line survey was distributed via multiple methods, such as at conferences and forums and through emails, newsletters, and social media. Respondents had the link from different channels, but they equally accessed to the same on-line survey. Each respondent was requested to answer two stated preference questions. The final dataset includes 549 respondents from US who responded in full. The survey details are presented in \citet{Le2017}.

\subsubsection{Data descriptions}

\noindent The collected data was compared to the 2014 US census data (https://www.census.gov/data/tables/2014/demo/age-and-sex/2014-age-sex-composition .html). The share of male respondents was a bit lower than in the US data. Other statistics closely reflect the corresponding census data. The sample description is presented in Table \ref{tab4.Socio_demogr}. \par 

\begin{table}
	\centering
	\caption{Socio-demographic characteristics}
	\scriptsize
	\label{tab4.Socio_demogr}               
	\begin{tabular}{p{10cm} p{4cm}}
		\hline
		\textbf{Attributes}	 & \textbf{Mean/ Standard Deviation or Distribution*}\\ [0.5ex] 		
		\hline
		Survey time & Jan 2017 - Apr 2017\\
		Survey location & US \\
		Total respondents & 549 \\ \hline
		Age            & 36.06/11.06                                                                                                   \\ \hline
		Gender: Male/ Female       & 45.50/ 54.50   \\ \hline
		
		Marital status: Single/ Married/ Others.                                                                                                                                                                                                                                                                   & \textbf{45.00}/ \textbf{44.80}/ 10.20                                                                                                                                                                                                                                                                                                                                                     \\ \hline
		Number of children.                                                                                   & 0.94/ 1.25                                                                                                                                                                                                                                           \\ \hline
		{Final academic degree: Some high school\textasciicircum 1/ High school diploma\textasciicircum 2/ Technical college degree\textasciicircum 3/ College degree\textasciicircum 4/ Post-graduate degree\textasciicircum 5/ I prefer not to answer\textasciicircum 6}                                                                                                                                     & {0.40\textasciicircum 1/ 12.90\textasciicircum 2/ 8.60\textasciicircum 3/ \textbf{48.50}\textasciicircum 4/ 29.00\textasciicircum 5/ 0.70\textasciicircum 6}                                                                                \\ \hline
		Annual income (\$1,000)                                                                                                                                                                                              & 48.71/ 36.00                                                                                                                                                                                                     \\ \hline
	\end{tabular}
\end{table}

\begin{figure} 
	\centering
	\begin{minipage}{.5\textwidth}
		\centering
		\includegraphics[width= 1 \textwidth]{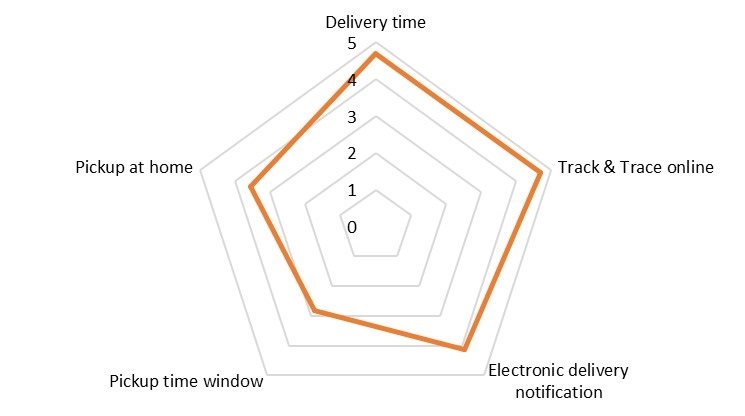}
		\caption{Satisfaction levels to past delivery \\services (normalize into five levels)}
		\label{fig4.Satisfiactions}	
	\end{minipage}%
	\begin{minipage}{.5\textwidth}
		\centering
		\includegraphics[width= 1 \textwidth]{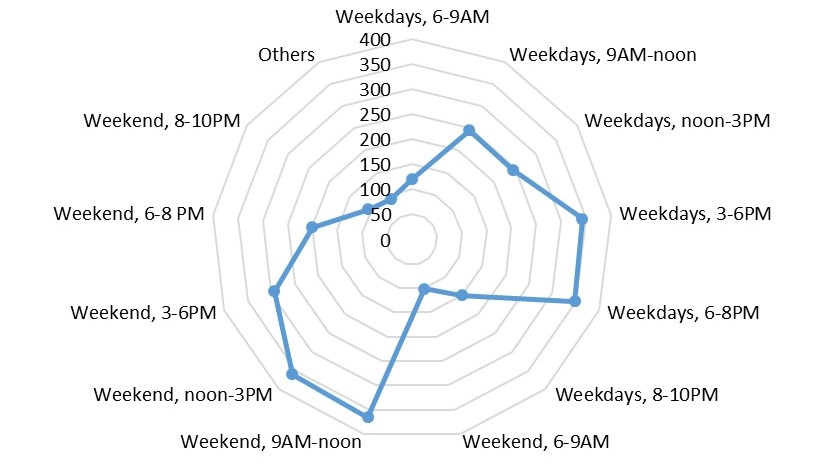}
		\caption{Preference on delivery time (multiple choices) (unit: respondents)}
		\label{fig4.Deliv_time}	
	\end{minipage}
\end{figure}

Figure \ref{fig4.Satisfiactions} illustrates the satisfaction-levels of survey respondents who had previous experiences with shipping service providers. The majority of respondents were happy with the delivery time, online tracking info, and electronic delivery notification services. Nevertheless, some respondents expressed less satisfaction with the pickup at home or pickup time-window services.

Figure \ref{fig4.Deliv_time} shows delivery time preferences. A larger portion of respondents preferred goods being delivered from 3PM to 8PM during weekdays, or from 9AM to 6PM during weekends. Weekdays after 8PM, weekends after 6PM, or every day before 9AM were less preferred options. The information is valuable for \<CS> companies setting up business operations strategies for the different time of the day or the day of the week. \par

\begin{figure} 
	\centering
	\includegraphics[width=0.8\textwidth]{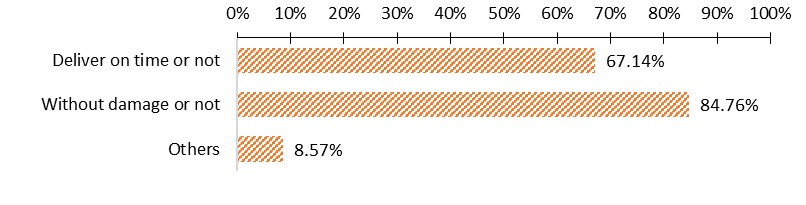}
	\caption{ Senders' concerns when their packages are delivered by \<CS> couriers (multiple choices)}
	\label{fig4.Concerns}	
\end{figure}

The main concerns respondents have about their packages being delivered by \<CS> couriers are exhibited in Figure \ref{fig4.Concerns}. Predominantly, respondents worried about their packages being delivered in good condition and on time. The findings are consistent with the study of \citet{cavalcante2013shipper} in traditional logistics literature. \par

\par

\begin{figure} 
	\centering
	\includegraphics[width=0.8\textwidth]{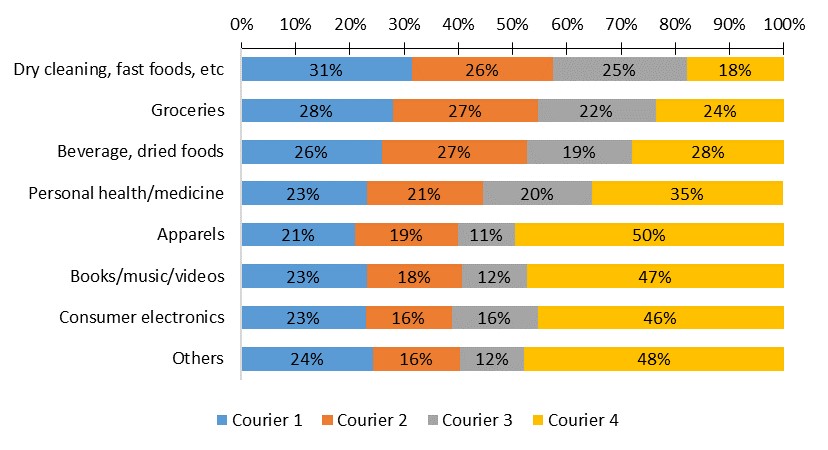}
	\caption{Senders select couriers for sending different types of products (Stated Preference)}
	\label{fig4.Choices}	
\end{figure}

As can be seen in Figure \ref{fig4.Choices}, respondents were found more likely to select couriers 1-3 for sending dry cleaning, fast foods, groceries, beverages, and dried foods. Those products may need to be delivered in a short time, which is the advantage of couriers 1-3 even though the delivery costs are higher than that of courier 4. In contrast, apparel, books, music, videos, consumer electronics, or other goods were highly preferred to be delivered by courier 4., which offered the lowest cost. \par

\par

\begin{figure}
	\centering
	\includegraphics[width=0.9 \textwidth]{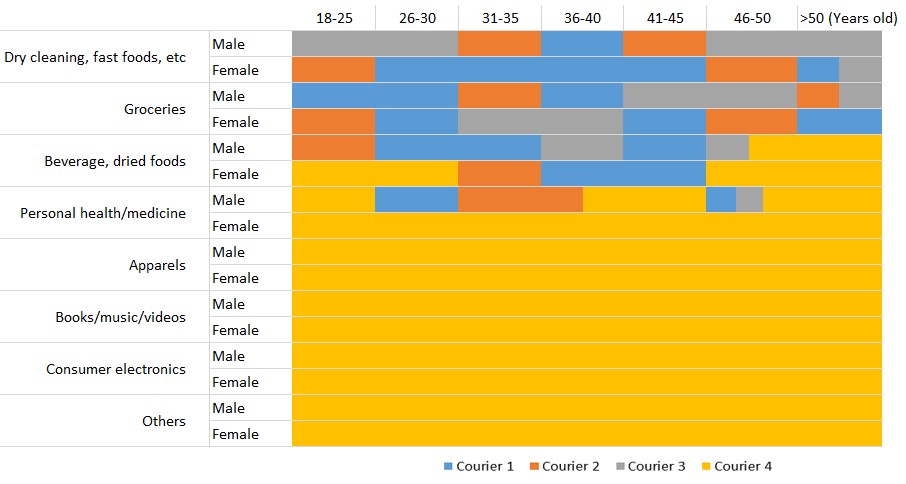}
	\caption{Courier choice's classification by genders, ages, and products}
	\label{fig4.GenAgeChoice}	
\end{figure}

\citet{Le2017} have provided a descriptive analysis and trends of the stakeholders from both demand and supply sides. The paper has also investigated how gender and age figure in to the respondents' willingness to work as \<CS> couriers. In Figure \ref{fig4.GenAgeChoice}, we extend the knowledge of \<CS> stakeholders (i.e., senders) by displaying the most-chosen courier service for different product categories, classified by gender and age-range. As can be observed, dry cleaning, fast foods, groceries, beverages, and dried foods are more likely to be sent via couriers 1-3 by both men and women of any age. Meanwhile, apparel, books, music, videos, consumer electronics, and other products are more likely to be sent via courier 4, regardless of gender or age. Personal health and medicine shipments seem to be a potential market segment for both \<CS> and \<TLCs>.

\par

The collected data was then analyzed to figure out the relative importance factors had on influencing the senders' choice of couriers and their WTP for the desired service. The models we used for analysis are presented in the next section. \par

\subsection{Statistical Modeling Approaches}
\noindent This section will present both the mixed logit and random regret formulations as modeling approaches for this research.

\subsubsection{Random utility maximization (RUM) model}
\noindent The mixed logit model, which is sometimes called a random-parameter multinomial logit model, is a popular approach used to capture variables' heterogeneities across observations. The utility function for an alternative $i$ of observation $n$ is defined in Equation \ref{eq.Util4}.
\begin{equation}\label{eq.Util4}
U_{in} =V_{in} +\xi_{in}
\end{equation}

\noindent where the observable utility is denoted as $ V_{in} $; the disturbance term that is represented as $ \xi $ follows the IID $ Gumbel $ distribution. The probability that an observation $n$ selecting an alternative $i$ from a given choice set of $J$ alternatives is then computed by Equation \ref{eq.MixP4}.
\begin{equation}\label{eq.MixP4}
P_{n}(i) =\int_{X} \frac{exp\big(\beta_{i}X_{in}\big)}{\sum_{j=1}^{J}exp\big(\beta_{j}X_{jn}\big)} {f(\beta|\varphi)} {d\beta}
\end{equation}

\noindent where $\beta_{i}$ is a vector of estimable parameters for the alternative $i$; $X_{in}$ is a vector of variables for the alternative $i$ of the observation $n$; $f(\beta|\varphi)$ is a density function.

\noindent The mixed logit model is usually estimated by using utility maximization methods. Therefore, it is also known as the random utility maximization (RUM) model.

\subsubsection{Random regret minimization (RRM) model}

\noindent The regret model \citep{chorus2008random}, however, is based on choice theory, which holds that a decision-maker will try to select the best option from the choice set in order to avoid a regret experience. Otherwise, the decision-maker will be in regret as the non-selected option is outweighed by the selected one. Assume any alternative in the choice set of $J$ alternatives has $K$ attributes ($k=1..K$). The regret of choosing an alternative $i$ instead of an alternative $j$ in terms of attribute $k$ is defined in Equation \ref{eq.regretfun4}.
\begin{equation}\label{eq.regretfun4}
R_{i\leftrightarrow j}(k) = ln\big[1+exp\big(\beta_{k}(x_{jk}-x_{ik})\big)\big]
\end{equation}

\noindent where $\beta_{k}$ is an estimable parameter for the attribute $k$; $x_{jk}$ and $x_{ik}$ are the attributes $k$ of the $j$ and $i$ alternatives.

\noindent The regret function for an alternative $i$ with $K$ attributes is displayed in Equation \ref{eq.regetfun24}.
\begin{equation}\label{eq.regetfun24}
R_{i} = \sum_{i\neq j} \sum_{k=1..K} ln\big[1+exp\big(\beta_{k}(x_{jk}-x_{ik})\big)\big]
\end{equation}

\noindent Since the computation for minimizing a function is equal to maximizing the negative of that function, the probability of choosing an $i$ alternative is then computed by Equation \ref{eq.probreget4}.
\begin{equation}\label{eq.probreget4}
P_{i} =\frac{exp\big(-R_{i}\big)}{\sum_{j=1}^{J}exp\big(-R_{j}\big)}
\end{equation}

\subsubsection{Willingness-to-pay estimation method}

\noindent In this study, \<WTP> will be computed by traditional methods \citep{small2012valuation}. All variables relating to \<CS> services, such as shipping cost, courier's reputation, tracking and tracing ability, electronic delivery notification, personalization for delivery time window/location, and tipping, will be estimated for \<WTP> by Equation \ref{eq.WTP4}.

\begin{equation}\label{eq.WTP4}
WTP= \left|\frac{\beta_{time}}{\beta_{i}}\right| (\$/h)	  
\end{equation}
\noindent where $\beta_{time}$ is an estimable parameter of the delivery time. $\beta_{i}$ is an estimable parameter of the $i$ variable (i.e. the attribute $i$ of the \<CS> service).

\section{Estimation results}

\noindent For model estimation, we use stated preference data which includes eight product (PD) categories. PD1 includes dry cleaning, fast food, and similar products which typically require quick delivery. PD2 includes groceries. PD3 contains beverages and dried foods. PD4 contains personal health and medicine products. PD5 contains apparel products. PD6 contains books, music, and videos products. PD7 contains consumer electronics products. PD8 contains other products. In addition, the choice-set has four alternatives (i.e. couriers 1-4). The variables of models include alternatives' attributes and respondents' socio-demographic characteristics.

Using the aforementioned data, two models, namely RUM and RRM, are used to estimate factors which are influence courier choices for sending each PD category. Overall, the two models' goodness-of-fit values are found comparable, even though those of RUM models are a bit larger than those of the corresponding RRM models. The similarity in performances of the two models is also displayed in some other transportation studies, such as \citet{hensher2013random,chorus2012random}.

In those RUM and RRM models, the shipping costs were estimated as a normally-distributed random parameter while other shipping attributes were computed as non-random parameters. The socio-demographic variables were also examined in the models. All variables were found to be significant at 90\% level or more, except some constant terms. The estimation results are summarized in Table \ref{tab4.RumRrmResults}. The following subsections will discuss insights from the estimated parameters.

\newgeometry{bottom=2cm, left=2cm, right=2cm}
\begin{landscape}
	\begin{ThreePartTable}
		\centering
		\tiny
		\makegapedcells
		\captionsetup{font=scriptsize}
		
		\begin{longtable} {llll|ll|ll|ll|ll|ll|ll|ll}
			\caption{Estimation results}	
			\label{tab4.RumRrmResults} \\
			\hline
			
			\multicolumn{2}{c}{\multirow{2}{*}{\thead{\textbf{Variables}}}}                                                                                                 & \multicolumn{2}{c}{\thead{\textbf{PD1}}}                                    & \multicolumn{2}{c}{\thead{\textbf{PD2}}}                                    & \multicolumn{2}{c}{\thead{\textbf{PD3}}}                                    & \multicolumn{2}{c}{\thead{\textbf{PD4}}}                                    & \multicolumn{2}{c}{\thead{\textbf{PD5}}}                                    & \multicolumn{2}{c}{\thead{\textbf{PD6}}}                                    & \multicolumn{2}{c}{\thead{\textbf{PD7}}}                                    & \multicolumn{2}{c}{\thead{\textbf{PD8}}}                                    \\ \cline{3-18}
			\multicolumn{2}{c}{}                                                                                                                                    & \multicolumn{1}{c}{\thead{\textbf{RUM}}} & \multicolumn{1}{c}{\thead{\textbf{RRM}}} & \multicolumn{1}{c}{\thead{\textbf{RUM}}} & \multicolumn{1}{c}{\thead{\textbf{RRM}}} & \multicolumn{1}{c}{\thead{\textbf{RUM}}} & \multicolumn{1}{c}{\thead{\textbf{RRM}}} & \multicolumn{1}{c}{\thead{\textbf{RUM}}} & \multicolumn{1}{c}{\thead{\textbf{RRM}}} & \multicolumn{1}{c}{\thead{\textbf{RUM}}} & \multicolumn{1}{c}{\thead{\textbf{RRM}}} & \multicolumn{1}{c}{\thead{\textbf{RUM}}} & \multicolumn{1}{c}{\thead{\textbf{RRM}}} & \multicolumn{1}{c}{\thead{\textbf{RUM}}} & \multicolumn{1}{c}{\thead{\textbf{RRM}}} & \multicolumn{1}{c}{\thead{\textbf{RUM}}} & \multicolumn{1}{c}{\thead{\textbf{RRM}}}  \\ \hline

			\endfirsthead
			
			\caption{Estimation results (cont.)} \\
			\hline 
			\multicolumn{2}{c}{\multirow{2}{*}{\thead{\textbf{Variables}}}}                                                                                                 & \multicolumn{2}{c}{\thead{\textbf{PD1}}}                                    & \multicolumn{2}{c}{\thead{\textbf{PD2}}}                                    & \multicolumn{2}{c}{\thead{\textbf{PD3}}}                                    & \multicolumn{2}{c}{\thead{\textbf{PD4}}}                                    & \multicolumn{2}{c}{\thead{\textbf{PD5}}}                                    & \multicolumn{2}{c}{\thead{\textbf{PD6}}}                                    & \multicolumn{2}{c}{\thead{\textbf{PD7}}}                                    & \multicolumn{2}{c}{\thead{\textbf{PD8}}}                                    \\
			\multicolumn{2}{c}{}                                                                                                                                    & \multicolumn{1}{c}{\thead{\textbf{RUM}}} & \multicolumn{1}{c}{\thead{\textbf{RRM}}} & \multicolumn{1}{c}{\thead{\textbf{RUM}}} & \multicolumn{1}{c}{\thead{\textbf{RRM}}} & \multicolumn{1}{c}{\thead{\textbf{RUM}}} & \multicolumn{1}{c}{\thead{\textbf{RRM}}} & \multicolumn{1}{c}{\thead{\textbf{RUM}}} & \multicolumn{1}{c}{\thead{\textbf{RRM}}} & \multicolumn{1}{c}{\thead{\textbf{RUM}}} & \multicolumn{1}{c}{\thead{\textbf{RRM}}} & \multicolumn{1}{c}{\thead{\textbf{RUM}}} & \multicolumn{1}{c}{\thead{\textbf{RRM}}} & \multicolumn{1}{c}{\thead{\textbf{RUM}}} & \multicolumn{1}{c}{\thead{\textbf{RRM}}} & \multicolumn{1}{c}{\thead{\textbf{RUM}}} & \multicolumn{1}{c}{\thead{\textbf{RRM}}}  \\ \hline
			
			\endhead
			\multicolumn{18}{r}{\textit{continuation on the next page}}
			\endfoot
			\endlastfoot
			
			\multicolumn{18}{l}{\textbf{Random parameters}} \\ 		
			\multicolumn{2}{l}{Shipping cost - mean}                                                                                                                   & (--)**                            & (--)**                            & (--)**                            & (--)**                            & (--)**                            & (--)**                            & (--)**                            & (--)**                            & (--)**                            & (--)**                            & (--)**                            & (--)**                            & (--)**                            & (--)**                            & (--)**                            & (--)**                             \\
			\multicolumn{2}{l}{Shipping cost - std}                                                                                                                   & (+)**                            & (+)**                            & (+)**                            & (+)**                            & (+)**                            & (+)**                            & (+)**                            & (+)**                            & (+)*                             & (+)**                            & (+)*                             & (+)*                             & (+)**                            & (+)**                            & (+)**                            & (+)**                             \\
			\hline
			\multicolumn{18}{l}{\textbf{Non-random parameters}}  \\ 
			\multirow{11}{*}{\rotatebox[origin=c]{90}{Shipping attributes}} & Delivery time                                                                                                   & (--)**                            & (--)**                            & (--)**                            & (--)**                            & (--)**                            & (--)**                            & (--)**                            & (--)**                            & (--)*                             & (--)*                             & -                                & -                                & (--)**                            & (--)**                            & (--)*                             & (--)**                             \\
			& Courier's reputation/ ranking                                                                                   & (+)**                            & (+)**                            & (+)**                            & (+)**                            & (+)**                            & (+)**                            & (+)**                            & (+)**                            & (+)**                            & (+)**                            & (+)**                            & (+)**                            & (+)**                            & (+)**                            & (+)**                            & (+)**                            \\
			& Tracking and tracing ability                                                                                    & (+)**                            & (+)**                            & (+)**                            & (+)*                             & (+)**                            & (+)**                            & (+)**                            & (+)**                            & (+)**                            & (+)**                            & (+)**                            & (+)**                            & (+)**                            & (+)**                            & (+)**                            & (+)**                             \\
			& Electronic delivery notification                                                                                & (+)**                            & (+)**                            & (+)**                            & (+)*                             & (+)**                            & (+)**                            & (+)**                            & (+)**                            & (+)**                            & (+)**                            & (+)**                            & (+)**                            & (+)**                            & (+)**                            & (+)**                            & (+)**                            \\
			& \begin{tabular}[c]{@{}l@{}}Personalization for \\delivery time window\end{tabular}                            & (+)**                            & (+)**                            & (+)**                            & (+)**                            & (+)**                            & (+)**                            & (+)**                            & (+)**                            & -                                & -                                & -                                & -                                & -                                & (+)**                            & -                                & (+)**                            \\
			& \begin{tabular}[c]{@{}l@{}}Personalization for \\location of delivery (not \\home or pickup point)\end{tabular} & (+)**                            & -                                & -                                & -                                & (--)**                            & (--)**                            & (--)**                            & (--)**                            & (--)**                            & (--)**                            & (--)**                            & (--)**                            & (--)**                            & (--)**                            & (--)**                            & -                                \\
			& \begin{tabular}[c]{@{}l@{}}Willingness to tip (for couriers 1, \\2, and 3)\end{tabular}                                                                                             & -                            & -                                & -                                & -                                & -                                & -                                & -                                & -                                & (+)*                             & -                                & (+)**                            & -                                & (+)**                            & -                                & -                                & -                                \\
			& Willingness to tip (for courier 1)                                                                              & (+)**                                & (+)**                            & (+)**                            & (+)**                            & (+)**                            & (+)**                            & (+)**                            & (+)**                            & -                                & -                                & -                                & -                                & -                                & -                                & -                                & -                                \\
			& Willingness to tip (for courier 2)                                                                              & (+)**                                & (+)**                            & (+)**                            & (+)**                            & (+)**                            & (+)**                            & (+)**                            & (+)**                            & -                                & -                                & -                                & -                                & -                                & -                                & -                                & -                                \\
			& Willingness to tip (for courier 3)                                                                              & (+)**                                & (+)**                            & (+)**                            & (+)**                            & (+)**                            & (+)**                            & (+)**                            & (+)**                            & -                                & -                                & -                                & -                                & -                                & -                                & -                                & -                                \\
			& \begin{tabular}[c]{@{}l@{}}Willingness to tip (for courier 4)\end{tabular}                                                                              & 0                                & 0                            & 0                                & 0                            & 0                                & 0                            & 0                                & 0                            & 0                                & -                                & 0                                & -                                & 0                                & -                                & -                                & -                                \\
			& \begin{tabular}[c]{@{}l@{}}Concerns on packages of being \\damaged\end{tabular}                                                                           & -                                & -                                & -                                & -                                & -                                & -                                & -                                & -                                & -                                & -                                & (+)*                             & -                                & -                                & -                                & -                                & -                               \\ \hline \pagebreak
			\multirow{11}{*}{\rotatebox[origin=c]{90}{Social demographic}}  & 21-25 years old                                                                                                 & -                                & -                                & (--)*                             & -                                & (+)**                            & -                                & -                                & -                                & -                                & -                                & -                                & -                                & -                                & -                                & -                                & -                                \\
			& 26-30  years old male                                                                                           & (+)**                            & -                                & (+)**                            & -                                & (+)**                            & -                                & -                                & -                                & -                                & -                                & -                                & -                                & -                                & -                                & -                                & -                                \\
			& 36-40 years old                                                                                                 & -                                & -                                & (+)**                            & -                                & (+)**                            & -                                & -                                & -                                & -                                & -                                & -                                & -                                & -                                & -                                & -                                & -                                \\
			& \begin{tabular}[c]{@{}l@{}}Personal annual income of \\\$100 -\$150 million\end{tabular}                                                                    & (+)*                             & -                                & -                                & -                                & -                                & -                                & -                                & -                                & -                                & -                                & -                                & -                                & -                                & -                                & -                                & -                                 \\
			& Married                                                                                                         & (+)*                             & -                                & -                                & -                                & -                                & -                                & -                                & -                                & -                                & -                                & -                                & -                                & -                                & -                                & -                                & -                                \\
			& \begin{tabular}[c]{@{}l@{}}Married and living with \\people (more than 64 years old) \end{tabular}                                                           & -                                & -                                &                                  & -                                & -                                & -                                & (+)**                            & -                                & (+)*                             & -                                & -                                & -                                & -                                & -                                & -                                & -                                \\
			& \begin{tabular}[c]{@{}l@{}}Living with people (less than \\18 years old)\end{tabular}                                                                       & -                                & -                                & -                                & -                                & -                                & -                                & (+)*                             & -                                & -                                & -                                & (+)**                            & -                                & (+)**                            & -                                & (+)**                            & -                                \\
			& \begin{tabular}[c]{@{}l@{}}Living with people (from 18 \\to 64 years old)\end{tabular}                                                                      & -                                & -                                & -                                & -                                & (+)**                            & -                                & -                                & -                                & (+)**                            & -                                & -                                & -                                & -                                & -                                & -                                & -                                \\
			& Full-time employees                                                                                             & (--)*                             & -                                & -                                & -                                &                                  & -                                & -                                & -                                & -                                & -                                & -                                & -                                & -                                & -                                & -                                & -                                \\
			& Full- and part-time employees                                                                                   & -                                & -                                & (+)**                            & -                                & (+)**                            & -                                & -                                & -                                & -                                & -                                & -                                & -                                & -                                & -                                & -                                & -                                \\
			& Full-time male employees                                                                                        & -                                & -                                & -                                & -                                & -                                & -                                & (+)*                             & -                                & (+)**                            & -                                & -                                & -                                & (+)**                            & -                                & (+)**                            & -                                \\ \hline
			\multirow{4}{*}{\rotatebox[origin=c]{90}{Constants}}            & Constant - courier 1                                                                                            & (--)**                            & (--)**                            & (--)**                            & (--)**                            & (--)**                            & (--)**                            & (--)**                            & (--)**                            & (--)**                            & (--)*                             & (--)**                            & (--)                              & (--)**                            & (--)**                            & (--)**                            & (--)**                            \\
			& Constant - courier 2                                                                                            & (--)**                            & (--)**                            & (--)**                            & (--)**                            & (--)**                            & (--)**                            & (--)**                            & (--)**                            & (--)**                            & (--)                              & (--)**                            & (--)                              & (--)**                            & (--)**                            & (--)**                            & (--)**                            \\
			& Constant - courier 3                                                                                            & (--)**                            & (--)**                            & (--)**                            & (--)**                            & (--)**                            & (--)**                            & (--)**                            & (--)**                            & (--)                              & (--)                              & (--)                              & (--)*                             & (--)                              & (+)                              & (--)                              & (--)                              \\
			& Constant - courier 4 (base)                                                                                     & 0                                & 0                                & 0                                & 0                                & 0                                & 0                                & 0                                & 0                                & 0                                & 0                                & 0                                & 0                                & 0                                & 0                                & 0                                & 0                                \\
			\hline \hline
			\multicolumn{18}{l}{\textbf{Model fits}}                                                                                                                                                                                                                                                                                                                                                                                                                                                                                                                                                                                                                                                                                                \\ 
			\multicolumn{2}{l}{Observation numbers}                                                                                                                 & 1098                             & 1098                             & 1098                             & 1098                             & 1098                             & 1098                             & 1098                             & 1098                             & 1098                             & 1098                             & 1098                             & 1098                             & 1098                             & 1098                             & 1098                             & 1098                             \\
			\multicolumn{2}{l}{Null log likelihood}                                                                                                                 & -1522                            & -1522                            & 1522                             & -1522                            & -1522                            & -1522                            & -1522                            & -1522                            & -1522                            & -1522                            & -1522                            & -1522                            & -1522                            & -1522                            & -1522                            & -1522                            \\
			\multicolumn{2}{l}{Final log likelihood}                                                                                                                & -1303                            & -1313                            & -1337                            & -1347                            & -1364                            & -1379                            & -1321                            & -1329                            & -1137                            & -1148                            & -1149                            & -1158                            & -1175                            & -1188                            & -1159                            & -1177                            \\
			\multicolumn{2}{l}{Rho square}                                                                                                                          & 0.144                            & 0.137                            & 0.122                            & 0.115                            & 0.104                            & 0.094                            & 0.132                            & 0.127                            & 0.253                            & 0.246                            & 0.245                            & 0.239                            & 0.228                            & 0.220                            & 0.239                            & 0.227                            \\
			\multicolumn{2}{l}{Adj Rho square}                                                                                                                      & 0.139                            & 0.134                            & 0.117                            & 0.112                            & 0.098                            & 0.090                            & 0.127                            & 0.123                            & 0.250                            & 0.244                            & 0.243                            & 0.237                            & 0.225                            & 0.217                            & 0.236                            & 0.224                            \\ \hline
			
		\end{longtable}
		\begin{compactenum}
			\item `PD1': Dry cleaning, fast foods, lunches, dinners, birthday cakes, etc (immediate delivery); `PD2': Groceries; PD3: Beverage/ dried foods; `PD4': Personal health/ medicine; `PD5': Apparels; PD6: Books/ Music/ Videos; `PD7': Consumer electronics; `PD8': Others.
			\item ** and *: Significance at 5\% and 10\% levels, respectively.
			\item (+) and (--): Positive and negative parameters.
			\item -: Variable is not included in the model.
		\end{compactenum}	
	\end{ThreePartTable}
\end{landscape}

\restoregeometry

\subsection{Attributes related to shipping services}

\noindent The significant parameters of various \<CS> service variables indicate the remarkable roles that service characteristics plays in the senders' choices of couriers. Together with the shipping cost and delivery time, other shipping attributes related to personalized services and services' quality are also noteworthy because they had a significant impact on senders' choices.

Shipping costs were found to be statistically significant and randomly distributed. The negatively estimated parameters suggest that senders prefer to send packages by a courier who offers a lower cost. This finding is consistent with the common knowledge on consumers' behaviors. While senders' behaviors are remarkably influenced by shipping cost, they also vary from sender to sender. As such, the shipping costs' parameters are heterogeneous across senders following a normal distribution. Meanwhile, delivery time parameters are also negative and significant for all PDs, from PD1 to PD8 (except PD6). Senders prefer to have shipments delivered in a shorter time. These findings on the negativity of the two variables are expected and alike with other studies of \citet{train2008estimation,anderson2009demand,tussyadiah2016factors,shaheen2016casual,briffaz2016crowd,punel2017modeling}.  

``Courier's reputation (ranking),'' ``Tracking and tracing ability,'' and ``Electronic delivery notification'' parameters were found to be positively significant. Those are three notable variables which senders tend to take into account when sending packages. Reputation reflects a courier's performance and delivery quality, while tracking and tracing provides a package's location in real-time. The electronic delivery notification service informs senders of changes in activity associated to their shipments. As expected, those three service features profoundly influence the senders' choice of couriers. In reality, the finding on the couriers' performance is alike with the traditional logistics study of \citet{cavalcante2013shipper} and the \<CS> study of \citet{punel2017modeling}. As practical applications from findings of three remarkable variables, \<CS> companies should provide such services and maintain those in a high level to attract more senders.

Furthermore, personalized services, such as setting the delivery time window and the preferred location of delivery, were found to be significant. Allowing for a customizable time-window positively influences the choice to ship packages with a certain courier. As a side note, the personalized time window does not necessarily mean immediate delivery. Interestingly, the personalized location of delivery parameters are found to be positive in situations where senders request \<CS> services to deliver PD1 items (dry cleaning, fast food, etc.), but negative for other product categories. Senders shipping PD1 products prefer them to be delivered in a short time to places other than home or pickup points, such as offices for example. In contrast, on-line buyers typically expect other products to be delivered to home or be picked up at one of the carriers' pickup points. These findings on personalized services are supported by revealed preference (RP) data from a study of \citet{Le2017}. In that study, many respondents were not happy with personalized time windows and delivery location options. About 44\% of respondents said "the carrier did not offer a pickup time window", while approximately 30\% of respondents said either the pickup at home service was bad or that they did not know about the service. Moreover, about 34\% of respondents reported that "carriers offer pickup at home,''  but for some reason they "have never used the service". For those reasons, our findings and the supported RP data are helpful for logistics firms identifying services which could be improved.

Tipping, an appreciation given to couriers from customers in the form of additionally payment, have been found to have positive significance on the choice of some \<CS> couriers over some traditional logistics carriers. Respondents are willing to tip to maximize their utility for all products (except "Other" products--PD8). Interestingly, the willingness to tip in the RRM models reveals respondents are more likely to tip for some \<CS> couriers to get dry cleaning, fast food, beverages, dried foods, groceries, personal health, and medicine products delivered.

On the other hand, safety has been identified as one of the challenges for implementing \<CS> services \citep{rouges2014crowdsourcing,carbone2015carried}. Concerns about shipped items being damaged have been found to be significant for books, music, and videos products (PD6) in our study. \<CS> companies should consider different forms of insurance or product guarantee for customers in case of damaged products.

The findings are valuable for \<CS> companies to address and improve service components which can attract more customers and influence senders' choices. More discussion on the parameter sensitivity and WTP for each service component will follow in the next sections.

\subsection{Socio-demographic characteristics}

\noindent Senders may have diverse considerations that govern their choice of courier for sending different types of packages. The choices also vary across population groups. An individual's socio-demographic characteristics significantly influence their behavior in attempt to maximize utility. The estimated results reveal PD1, PD2, and PD3 are more likely being sent by married, 21-40 years old, and middle income people. Likewise, living with young people, adults, or elderly people have certain influences on behaviors of the senders of PD3 to PD8. Full- and part-time employees are found remarkably for sending all kind of shipment categories. In fact, ride-sharing studies have revealed that influencing factors include age, gender, income, education level, and employment status \citep{dias2017behavioral,kooti2017analyzing}. Moreover, \citet{ballare2018preliminary} have found that popular users of the \<CS> services are between 35-44 years old, whereas, \citet{punel2018} found young men and full-time workers are more likely to use \<CS> services. The two studies, however, do not specify users for different shipment classes. Collectively, the findings of some particular socio-demographic characteristics of \<CS> users are valuable for \<CS> companies in addressing potential market segments (i.e. shipment classes) and developing revenue models (i.e. pricing strategies).

\section{Senders' perceptions and elasticity analysis}

\noindent This section aims to provide additional insights for \<CS> companies to improve their business strategies and services by understanding senders' perceptions of their services along with their competitors' services. With that intention, results were computed for all products (i.e. PD1 to PD8) to reveal senders' behaviors. Accordingly, we decided to conduct analysis on both RUM and RRM models and compared absolute dissimilarities as well as percent differences. The results, insights, and discussions are presented in the following sub-sections.

\subsection{Willingness to pay}

\noindent It is essential to know how much senders are willing to pay for a better service, especially in the logistics market in which new \<CS> firms compete against \<TLCs>s. For each product category, even the RUM and RRM models' performances are similar but the parameters' sizes vary. Parameters' sizes are important since the implementation are generally based on these parameters' magnitudes. In terms of WTPs, for each PD, results from the RRM model are remarkably different from those of the RUM model because of its attribute- and alternative-defined estimations, naturally by the RRM function's formulation. The values which are computed from the RUM and RRM models can be considered as lower and upper bounds for each corresponding WTP. As such, \<CS> firms have WTP ranges for references. The result details are presented in Table \ref{tab4.WTP4}. 

Among seven PDs, respondents were willing to pay higher rates for shipping PD1, PD2, and PD3 and lower rates for PD5, PD7, and PD8, for any service. More specifically, the \<WTP> values for sending PD1, PD2, and PD3 are from about \$0.6 to \$5, whereas that range for PD5, PD7, and PD8 is from about \$0.1 to \$0.6. Meanwhile, the \<WTP> value for PD4 is around \$0.6 to \$1.1. Interestingly, respondents were willing to pay more to have PD1 and PD2 delivered by higher reputation driver-partners and \<CS> firms provide electronic delivery notification and tracking and tracing ability. This is possibly because PD1 and PD2 are mainly food items which respondents can only consume in a certain time window. The findings of \<WTP> are consistent with common knowledge. People are typically willing to pay more to receive personalized or better services. The WTP value of RUM model for PD1 is similar to the finding in the study of \citet{punel2017modeling}. Other WTP values, however, could not be verified since this is the first study on this topic. Future studies should obtain operational data from successful businesses for a reliable validation. 

\begin{figure} [ht]
	\centering
	\includegraphics[width=0.8 \textwidth]{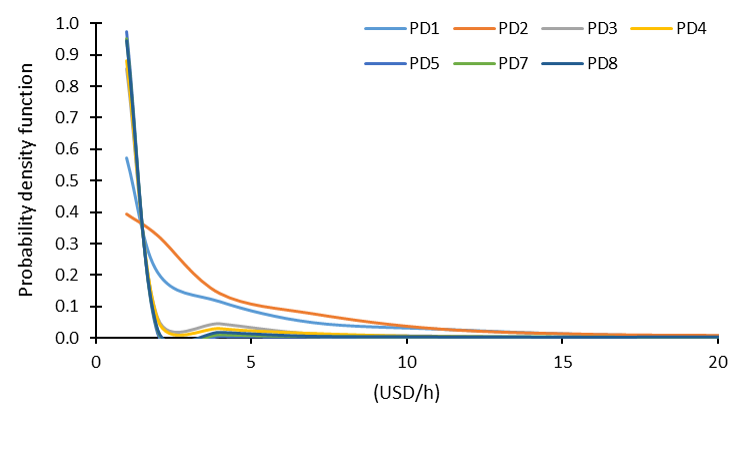}
	\caption{Probability density functions of the WTP for a delivery service for different products}
	\label{fig.WTPfun4}	
\end{figure}

In addition, the probability distribution functions of the senders' WTPs for a delivery service for the seven PDs are displayed in Figure \ref{fig.WTPfun4}. As can be seen, a significant portion of respondents are likely to pay about \$2 to \$10 for delivering PD1 and PD2. Whereas, for shipping PD5, PD7, and PD8, majority of respondents are likely to pay around \$2 to \$2.5. For delivering PD3 and PD4, high possibility that respondents pay approximately \$2 to \$5. Collectively, the WTP values and probability distributions are significant inputs for \<CS> firms to make pricing decisions.

\par

\newgeometry{bottom=1.8cm, left=2cm, right=2cm}
\begin{landscape}
	\begin{ThreePartTable}
		\centering
		\scriptsize
		\captionsetup{font=scriptsize}
		\begin{longtable}{llll|lll|lll|lll|lll|lll|lll}
			\caption{WTP estimation results}	
			\label{tab4.WTP4} \\
			\hline
			\multicolumn{1}{c}{\multirow{2}{*}{\textbf{WTP for}}} & \multicolumn{3}{c}{\textbf{PD1}} & \multicolumn{3}{c}{\textbf{PD2}} & \multicolumn{3}{c}{\textbf{PD3}} & \multicolumn{3}{c}{\textbf{PD4}} & \multicolumn{3}{c}{\textbf{PD5}} & \multicolumn{3}{c}{\textbf{PD7}} & \multicolumn{3}{c}{\textbf{PD8}} \\ \cline{2-22} 
			\multicolumn{1}{c}{}                          & RUM   & RRM   & Ratio & RUM   & RRM   & Ratio & RUM   & RRM   & Ratio & RUM   & RRM   & Ratio & RUM   & RRM   & Ratio & RUM   & RRM   & Ratio & RUM   & RRM   & Ratio \\ \hline
			Delivery service                                  & 2.097 & 3.065 & 1.462   & 1.848 & 4.741 & 2.566   & 0.586 & 3.738 & 6.380   & 0.528 & 1.020 & 1.931   & 0.130 & 0.286 & 2.212   & 0.101 & 0.363 & 3.599   & 0.097 & 0.594 & 6.129   \\
			Reputation                                    & 0.094 & 0.203 & 2.154   & 0.102 & 0.240 & 2.338   & 0.054 & 0.130 & 2.406   & 0.028 & 0.071 & 2.535   & 0.009 & 0.030 & 3.134   & 0.009 & 0.037 & 4.096   & 0.009 & 0.039 & 4.170   \\
			Tracking                                      & 0.090 & 0.272 & 3.035   & 0.106 & 0.276 & 2.605   & 0.056 & 0.144 & 2.582   & 0.039 & 0.110 & 2.845   & 0.009 & 0.030 & 3.320   & 0.008 & 0.038 & 4.718   & 0.007 & 0.037 & 4.971   \\
			E-notification                                & 0.129 & 0.213 & 1.653   & 0.123 & 0.293 & 2.385   & 0.044 & 0.102 & 2.310   & 0.026 & 0.068 & 2.662   & 0.006 & 0.018 & 3.204   & 0.006 & 0.027 & 4.528   & 0.006 & 0.043 & 7.751   \\
			P-Time                                        & 0.063 & 0.128 & 2.029   & 0.058 & 0.129 & 2.201   & 0.052 & 0.122 & 2.329   & 0.056 & 0.120 & 2.125   & -     & -     & -       & -     & 0.119 & -       & -     & 0.070 & -       \\
			P-Location                                    & 0.135 & -     & -       & -     & -     & -       & 0.063 & 0.141 & 2.244   & 0.033 & 0.090 & 2.773   & 0.008 & 0.025 & 3.253   & 0.009 & 0.044 & 4.857   & 0.007 & -     & -       \\
			Tip                                           & -     & -     & -       & -     & -     & -       & -     & -     & -       & -     & -     & -       & 0.028 & -     & -       & 0.025 & -     & -       & -     & -     & -       \\
			Tip1                                          & 0.097 & 0.210 & 2.173   & 0.127 & 0.259 & 2.044   & 0.051 & 0.122 & 2.400   & 0.076 & 0.171 & 2.239   & -     & -     & -       & -     & -     & -       & -     & -     & -       \\
			Tip2                                          & 0.081 & 0.191 & 2.346   & 0.071 & 0.138 & 1.935   & 0.043 & 0.107 & 2.488   & 0.043 & 0.095 & 2.240   & -     & -     & -       & -     & -     & -       & -     & -     & -       \\
			Tip3                                          & 0.065 & 0.130 & 1.992   & 0.062 & 0.112 & 1.800   & 0.030 & 0.060 & 2.010   & 0.029 & 0.057 & 1.966   & -     & -     & -       & -     & -     & -       & -     & -     & -       \\ \hline
		\end{longtable}
		
		\begin{compactenum}
			\item `PD1': Dry cleaning, fast foods, lunches, dinners, birthday cakes, etc (immediate delivery); `PD2': Groceries; PD3: Beverage/ dried foods; `PD4': Personal health/ medicine; `PD5': Apparels; `PD7': Consumer electronics; `PD8': Others.
			\item `Reputation': Courier's reputation/ranking; `Tracking': Tracking and tracing ability; `E-notification': Electronic delivery notification; `P-Time': Personalization for delivery time window; `P-Location': Personalization for location of delivery (not home or pickup point); `Tip': Average tip for the first three couriers (couriers1-3); `Tip1', `Tip2', `Tip3': Tip for couriers 1, 2, and 3, respectively.
			\item `Ratio': RRM/RUM.
		\end{compactenum}	
	\end{ThreePartTable}
\end{landscape}

\restoregeometry

\subsection{Elasticity analysis}
\noindent Elasticity analysis is commonly conducted to capture the changes in probability of selecting an alternative once changing in a certain variable while fixing other variables. In this research, direct elasticities were computed. Details of products' elasticities are presented in Table \ref{tab.PD1elasSum4} to Table \ref{tab.PD8elasSum4}. Moreover, the differences in percentages of attributes in RRM and RUM models for each product are illustrated in Figure \ref{fig.elas4} (Appendix B). The differences were computed by Equation \ref{eq.diffRRMRUM4}.

\begin{equation}\label{eq.diffRRMRUM4}
\Delta A_{i} (\%) =\frac{A_{i}(RRM) - A_{i}(RUM)}{A_{i}(RRM)}*100
\end{equation}
\noindent where $A_{i}$ is an attribute $i$ in the RRM or RUM models.

Overall, the shipping costs have the largest absolute elasticity magnitudes. The elasticities of personalized delivery time-window or tipping behaviors are remarkably large for PD1 to PD4, but not for others. Meanwhile, personalized delivery location elasticities are not significantly different for PD1 and PD2, but are significantly different for PD3 to PD7. As such, delivery times and delivery locations play important roles in selecting couriers for sending PD1 to PD4 and PD3 to PD7, respectively. Moreover, in RUM models for PD1, PD2, and PD3, the shipping cost, delivery time and reputation elasticities of courier 4 (i.e. C4) are remarkably larger than those of couriers 1-3. That means the shipping cost, delivery time, and reputation of the courier 4 should be significantly changed in order to make senders to select the courier 4. On the other hand, an opposite trend has been observed for shipping costs of couriers 1-3 versus those of courier 4 for PD5-PD8. \<CS> firms should lower a big amount of price to attract PD5-PD8 to be sent by their services. Collectively, \<CS> firms should provide various services to address different shipment categories.

The following notes are for Table \ref{tab.PD1elasSum4} to Table \ref{tab.PD8elasSum4}. 
\begin{compactenum}
	\item All ``\%'' columns were computed by Equation \ref{eq.diffRRMRUM4}.
	\item `C1', `C2', `C3', and `C4' represent for four alternatives, namely Courier 1, Courier 2, Courier 3, and Courier 4.
	\item `SCost': Shipping cost; `DTime': Delivery time; `Reputation': Courier's reputation/ranking; `Tracking': Tracking and tracing ability; `E-notification': Electronic delivery notification; `P-Time': Personalization for delivery time window; `P-Location': Personalization for location of delivery (not home or pickup point); `Tip': Willingness to tip.
\end{compactenum}

\begin{table} 
	\centering
	\scriptsize
	\caption{PD's elasticity summary}
	\begin{subtable} [c]{\textwidth} 
		\centering	
		\caption{PD1's elasticity summary}
		\label{tab.PD1elasSum4}
        \begin{tabular}{lrrr|rrr|rrr|rrr}
			\hline
			\multicolumn{1}{c}{\multirow{2}{*}{Variable}} & \multicolumn{3}{c}{C1}      & \multicolumn{3}{c}{C2}      & \multicolumn{3}{c}{C3}      & \multicolumn{3}{c}{C4}      \\ \cline{2-13} 
			\multicolumn{1}{c}{}                          & RUM & RRM & \%  & RUM & RRM & \%  & RUM & RRM & \%  & RUM & RRM & \%  \\ \hline
			SCost                                         & -0.632  & 0.362   & 274.703 & -0.625  & 0.380   & 264.412 & -0.050  & 0.538   & 109.268 & -1.003  & 0.326   & 407.666 \\
			DTime                                         & -0.506  & -1.713  & 70.494  & -0.232  & -0.236  & 1.653   & -0.067  & -0.091  & 26.484  & -1.600  & -0.911  & -75.628 \\
			Reputation                                    & 0.472   & 1.161   & 59.309  & 0.447   & 0.850   & 47.435  & 0.439   & 0.588   & 25.332  & 0.885   & 1.265   & 30.080  \\
			Tracking                                      & 0.236   & 0.412   & 42.746  & 0.152   & 0.217   & 29.949  & 0.110   & 0.130   & 15.355  & 0.178   & 0.168   & -6.444  \\
			E-notification                                & 0.151   & 0.494   & 69.528  & 0.121   & 0.323   & 62.697  & 0.095   & 0.248   & 61.724  & 0.046   & 0.281   & 83.476  \\
			P-Time                                        & 0.089   & 0.171   & 48.219  & 0.350   & 0.807   & 56.633  & 0.094   & 0.154   & 38.827  & 0.255   & 2.378   & 89.279  \\
			P-Location                                    & 0.156   & -       & -       & 0.062   & -       & -       & 0.141   & -       & -       & 0.113   & -       & -       \\
			Tip                                           & 0.503   & 1.176   & 57.211  & 0.622   & 1.170   & 46.862  & 0.714   & 1.822   & 60.799  & -       & -       & -       \\ \hline
		\end{tabular}
	\end{subtable}
	
	\begin{subtable} [c]{\textwidth} 
		\centering
		\scriptsize
		\caption{PD2's elasticity summary}
		\label{tab.PD2elasSum4}
		\begin{tabular}{lrrr|rrr|rrr|rrr}
			\hline
			\multicolumn{1}{c}{\multirow{2}{*}{Variable}} & \multicolumn{3}{c}{C1}      & \multicolumn{3}{c}{C2}      & \multicolumn{3}{c}{C3}      & \multicolumn{3}{c}{C4}      \\ \cline{2-13} 
			\multicolumn{1}{c}{}                          & RUM & RRM & \%  & RUM & RRM & \%  & RUM & RRM & \%  & RUM & RRM & \%  \\ \hline
			SCost          & -0.764 & -1.423 & 46.311 & -0.791 & -1.429 & 44.672 & -0.387 & -1.529 & 74.697 & -0.996 & -1.435 & 30.596  \\
			DTime          & -0.482 & -1.636 & 70.546 & -0.210 & -0.198 & -6.212 & -0.063 & -0.076 & 16.909 & -1.431 & -0.821 & -74.272 \\
			Reputation     & 0.427  & 0.936  & 54.412 & 0.394  & 0.591  & 33.299 & 0.385  & 0.429  & 10.266 & 0.721  & 0.834  & 13.606  \\
			Tracking       & 0.198  & 0.390  & 49.179 & 0.119  & 0.164  & 27.306 & 0.093  & 0.118  & 20.681 & 0.132  & 0.131  & -0.992  \\
			E-notification & 0.156  & 0.350  & 55.317 & 0.117  & 0.189  & 37.916 & 0.093  & 0.158  & 41.098 & 0.043  & 0.141  & 69.307  \\
			P-Time         & 0.106  & 0.178  & 40.641 & 0.357  & 0.626  & 42.990 & 0.102  & 0.138  & 25.654 & 0.241  & 0.999  & 75.906  \\
			Tip            & 0.387  & 0.977  & 60.444 & 0.662  & 1.229  & 46.148 & 0.705  & 1.711  & 58.830 & -      & -      & -      \\ \hline
		\end{tabular}
	\end{subtable}

	\begin{subtable} [c]{\textwidth} 
		\centering
		\scriptsize
		\caption{PD3's elasticity summary}
		\label{tab.PD3elasSum4}
		\begin{tabular}{lrrr|rrr|rrr|rrr}
			\hline
			\multicolumn{1}{c}{\multirow{2}{*}{Variable}} & \multicolumn{3}{c}{C1}      & \multicolumn{3}{c}{C2}      & \multicolumn{3}{c}{C3}      & \multicolumn{3}{c}{C4}      \\ \cline{2-13} 
			\multicolumn{1}{c}{}                          & RUM & RRM & \%  & RUM & RRM & \%  & RUM & RRM & \%  & RUM & RRM & \%  \\ \hline		
			SCost          & -1.366 & -0.656 & -108.216 & -1.458 & -0.661 & -120.645 & -0.369 & -0.840 & 56.054 & -1.886 & -0.555 & -239.564 \\
			DTime          & -0.275 & -0.560 & 50.964   & -0.126 & -0.141 & 10.100   & -0.039 & -0.058 & 33.505 & -0.789 & -0.624 & -26.342  \\
			Reputation     & 0.495  & 1.045  & 52.674   & 0.459  & 0.808  & 43.188   & 0.445  & 0.631  & 29.472 & 0.750  & 1.024  & 26.789   \\
			Tracking       & 0.231  & 0.445  & 48.146   & 0.135  & 0.222  & 39.225   & 0.107  & 0.237  & 54.588 & 0.139  & 0.164  & 15.633   \\
			E-notification & 0.275  & 0.633  & 56.624   & 0.200  & 0.325  & 38.478   & 0.165  & 0.396  & 58.264 & 0.065  & 0.118  & 45.478   \\
			P-Time         & 0.071  & 0.198  & 64.026   & 0.241  & 0.464  & 48.050   & 0.072  & 0.168  & 57.126 & 0.154  & 0.589  & 73.824   \\
			P-Location     & -0.206 & -0.483 & 57.279   & -0.077 & -0.128 & 39.860   & -0.184 & -0.472 & 60.991 & -0.132 & -0.540 & 75.607   \\
			Tip            & 0.514  & 1.072  & 52.070   & 0.583  & 0.986  & 40.884   & 0.785  & 2.375  & 66.954 & -      & -      & -  \\ \hline
		\end{tabular}
	\end{subtable}

	\begin{subtable} [c]{\textwidth} 
		\centering
		\scriptsize
		\caption{PD4's elasticity summary}
		\label{tab.PD4elasSum4}
		\begin{tabular}{lrrr|rrr|rrr|rrr}
			\hline
			\multicolumn{1}{c}{\multirow{2}{*}{Variable}} & \multicolumn{3}{c}{C1}      & \multicolumn{3}{c}{C2}      & \multicolumn{3}{c}{C3}      & \multicolumn{3}{c}{C4}      \\ \cline{2-13} 
			\multicolumn{1}{c}{}                          & RUM & RRM & \%  & RUM & RRM & \%  & RUM & RRM & \%  & RUM & RRM & \%  \\ \hline
			SCost          & -1.507 & -0.881 & -71.133 & -1.639 & -0.904 & -81.352 & -0.330 & -1.090 & 69.761 & -1.705 & -0.668 & -155.256 \\
			DTime          & -0.223 & -0.469 & 52.508  & -0.106 & -0.123 & 13.599  & -0.030 & -0.049 & 37.832 & -0.556 & -0.502 & -10.793  \\
			Reputation     & 0.773  & 1.722  & 55.121  & 0.739  & 1.450  & 49.024  & 0.670  & 0.952  & 29.650 & 0.995  & 1.452  & 31.499   \\
			Tracking       & 0.285  & 0.640  & 55.408  & 0.165  & 0.303  & 45.631  & 0.122  & 0.276  & 55.806 & 0.145  & 0.164  & 11.470   \\
			E-notification & 0.392  & 0.919  & 57.388  & 0.288  & 0.462  & 37.616  & 0.231  & 0.577  & 59.903 & 0.069  & 0.091  & 24.642   \\
			P-Time         & 0.056  & 0.246  & 77.289  & 0.191  & 0.487  & 60.730  & 0.054  & 0.161  & 66.168 & 0.101  & 0.387  & 73.999   \\
			P-Location     & -0.327 & -0.727 & 55.021  & -0.134 & -0.212 & 36.882  & -0.286 & -0.703 & 59.266 & -0.189 & -0.564 & 66.489   \\
			Tip            & 0.280  & 0.779  & 64.022  & 0.478  & 1.056  & 54.730  & 0.621  & 2.154  & 71.158 & -      & -      & -  		\\ \hline
		\end{tabular}
	\end{subtable}
\end{table}

\begin{table}
	\ContinuedFloat
	\caption{PD's elasticity summary (continued)}
	
	\begin{subtable} [c]{\textwidth} 
		\centering
		\scriptsize
		\caption{PD5's elasticity summary}
		\label{tab.PD5elasSum4}
		\begin{tabular}{lrrr|rrr|rrr|rrr}
			\hline
			\multicolumn{1}{c}{\multirow{2}{*}{Variable}} & \multicolumn{3}{c}{C1}      & \multicolumn{3}{c}{C2}      & \multicolumn{3}{c}{C3}      & \multicolumn{3}{c}{C4}      \\ \cline{2-13} 
			\multicolumn{1}{c}{}                          & RUM & RRM & \%  & RUM & RRM & \%  & RUM & RRM & \%  & RUM & RRM & \%  \\ \hline
			SCost          & -2.501 & -3.059 & 18.234 & -2.634 & -3.176 & 17.071 & -0.941 & -2.896 & 67.508 & -1.914 & -2.074 & 7.730  \\
			DTime          & -0.070 & -0.174 & 60.023 & -0.034 & -0.052 & 34.043 & -0.011 & -0.021 & 49.275 & -0.135 & -0.186 & 27.805 \\
			Reputation     & 0.755  & 2.581  & 70.764 & 0.768  & 2.471  & 68.924 & 0.696  & 1.035  & 32.737 & 0.697  & 1.037  & 32.771 \\
			Tracking       & 0.408  & 1.656  & 75.377 & 0.235  & 0.601  & 60.872 & 0.197  & 1.068  & 81.599 & 0.158  & 0.195  & 19.231 \\
			E-notification & 0.590  & 2.499  & 76.378 & 0.431  & 0.772  & 44.178 & 0.361  & 1.421  & 74.632 & 0.061  & 0.082  & 26.309 \\
			P-Location     & -0.457 & -1.842 & 75.195 & -0.191 & -0.322 & 40.807 & -0.434 & -2.137 & 79.696 & -0.213 & -0.388 & 45.063 \\
			Tip            & 0.210  & -      & -      & 0.211  & -      & -      & 0.194  & -      & -      & -      & -      & -     \\ \hline
		\end{tabular}
	\end{subtable}
	
	\begin{subtable} [c]{\textwidth} 
		\centering
		\scriptsize
		\caption{PD6's elasticity summary}
		\label{tab.PD6elasSum4}
		\begin{tabular}{lrrr|rrr|rrr|rrr}
			\hline
			\multicolumn{1}{c}{\multirow{2}{*}{Variable}} & \multicolumn{3}{c}{C1}      & \multicolumn{3}{c}{C2}      & \multicolumn{3}{c}{C3}      & \multicolumn{3}{c}{C4}      \\ \cline{2-13} 
			\multicolumn{1}{c}{}                          & RUM & RRM & \%  & RUM & RRM & \%  & RUM & RRM & \%  & RUM & RRM & \%  \\ \hline
			SCost          & -2.514 & -3.020 & 16.753 & -2.632 & -3.231 & 18.550 & -1.091 & -2.882 & 62.134 & -1.967 & -2.095 & 6.100  \\
			Reputation     & 0.648  & 2.293  & 71.757 & 0.686  & 0.946  & 27.516 & 0.587  & 0.933  & 37.056 & 0.612  & 0.941  & 35.002 \\
			Tracking       & 0.430  & 1.646  & 73.869 & 0.257  & 0.727  & 64.598 & 0.204  & 0.371  & 45.008 & 0.170  & 0.210  & 19.163 \\
			E-notification & 0.480  & 1.923  & 75.029 & 0.368  & 0.676  & 45.598 & 0.297  & 1.233  & 75.937 & 0.051  & 0.067  & 24.739 \\
			P-Location     & -0.405 & -1.639 & 75.284 & -0.169 & -0.295 & 42.804 & -0.381 & -1.810 & 78.931 & -0.192 & -0.311 & 38.360 \\
			Tip            & 0.219  & -      & -      & 0.229  & -      & -      & 0.202  & -      & -      & -      & -      & - \\ \hline
		\end{tabular}
	\end{subtable}
	
	\begin{subtable} [c]{\textwidth} 
		\centering
		\scriptsize
		\caption{PD7's elasticity summary}
		\label{tab.PD7elasSum4}
		\begin{tabular}{lrrr|rrr|rrr|rrr}
			\hline
			\multicolumn{1}{c}{\multirow{2}{*}{Variable}} & \multicolumn{3}{c}{C1}      & \multicolumn{3}{c}{C2}      & \multicolumn{3}{c}{C3}      & \multicolumn{3}{c}{C4}      \\ \cline{2-13} 
			\multicolumn{1}{c}{}                          & RUM & RRM & \%  & RUM & RRM & \%  & RUM & RRM & \%  & RUM & RRM & \%  \\ \hline
			SCost          & -2.138 & -2.186 & 2.195  & -2.300 & -2.342 & 1.793  & -0.143 & -2.333 & 93.867 & -1.844 & -1.476 & -24.975 \\
			DTime          & -0.080 & -0.265 & 69.785 & -0.041 & -0.081 & 49.444 & -0.011 & -0.031 & 65.176 & -0.159 & -0.280 & 43.082  \\
			Reputation     & 0.884  & 2.596  & 65.949 & 0.884  & 2.565  & 65.554 & 0.729  & 1.179  & 38.155 & 0.842  & 1.346  & 37.467  \\
			Tracking       & 0.510  & 1.526  & 66.575 & 0.299  & 0.737  & 59.449 & 0.203  & 0.730  & 72.247 & 0.204  & 0.243  & 16.097  \\
			E-notification & 0.617  & 1.776  & 65.240 & 0.478  & 0.923  & 48.163 & 0.360  & 1.387  & 74.008 & 0.068  & 0.086  & 20.468  \\
			P-Time         & -      & 0.219  & -      & -      & 0.437  & -      & -      & 0.160  & -      & -      & 0.133  & -       \\
			P-Location     & -0.431 & -1.135 & 62.035 & -0.191 & -0.325 & 41.316 & -0.371 & -1.388 & 73.258 & -0.212 & -0.409 & 48.103  \\
			Tip            & 0.299  & -      & -      & 0.305  & -      & -      & 0.254  & -      & -      & -      & -      & -  \\ \hline
		\end{tabular}
	\end{subtable}

	\begin{subtable} [c]{\textwidth} 
		\centering
		\scriptsize
		\caption{PD8's elasticity summary}
		\label{tab.PD8elasSum4}
		\begin{tabular}{lrrr|rrr|rrr|rrr}
			\hline
			\multicolumn{1}{c}{\multirow{2}{*}{Variable}} & \multicolumn{3}{c}{C1}      & \multicolumn{3}{c}{C2}      & \multicolumn{3}{c}{C3}      & \multicolumn{3}{c}{C4}      \\ \cline{2-13} 
			\multicolumn{1}{c}{}                          & RUM & RRM & \%  & RUM & RRM & \%  & RUM & RRM & \%  & RUM & RRM & \%  \\ \hline
			SCost          & -2.146 & -1.638 & -30.989 & -2.368 & -1.818 & -30.254 & -0.194 & -2.019 & 90.384 & -1.785 & -1.014 & -75.927 \\
			DTime          & -0.062 & -0.173 & 64.067  & -0.032 & -0.066 & 51.205  & -0.009 & -0.025 & 62.992 & -0.123 & -0.224 & 45.094  \\
			Reputation     & 0.677  & 1.705  & 60.285  & 0.711  & 2.266  & 68.618  & 0.625  & 0.953  & 34.393 & 0.640  & 0.963  & 33.551  \\
			Tracking       & 0.432  & 1.055  & 59.064  & 0.265  & 0.680  & 61.091  & 0.200  & 0.652  & 69.387 & 0.171  & 0.191  & 10.658  \\
			E-notification & 0.515  & 0.800  & 35.629  & 0.415  & 0.527  & 21.317  & 0.331  & 0.946  & 65.015 & 0.059  & 0.045  & -31.180 \\
			P-Time         & -      & 0.270  & -       & -      & 0.686  & -       & -      & 0.209  & -      & -      & 0.119  & -       \\
			P-Location     & -0.421 & -      & -       & -0.193 & -      & -       & -0.402 & -      & -      & -0.203 & -      & -       \\
			\hline
		\end{tabular}
	\end{subtable}
	
\end{table}

\section{Model validations}

\noindent The objective of this section is to validate the RUM and RRM models. As such, we collected the choice prediction values to evaluate the models' accuracies.

To begin, we randomly divided the dataset into five folds. For each fold, 80\% of the dataset (i.e. training set) are first used to estimate for the model parameters which are then fixed to estimate the model's performance for the second time with the remaining 20\% of the dataset (i.e. testing set). We ran each model five times. The choice prediction values of the five runs for each model are collected and compared to the actual choices in the corresponding 20\% of the dataset. In this paper, the choices are scalable. Therefore, the Mean Absolute Percentage Error (MAPE) is an appropriate method to evaluate for the accuracies of the prediction models. The MAPE value is computed by Equation \ref{eq.MAPEs4}.

\begin{equation}\label{eq.MAPEs4}
MAPE = \frac{100\%}{N} \sum_{i=1}^{N} \left| \frac{Y-\widehat{Y}}{Y} \right|
\end{equation}

\noindent in which $Y$ and $\widehat{Y}$ are the actual value (i.e. actual choice) and the predicted value (i.e. prediction choice). $N$ is the number of fitted points (e.g., $N$=5 in our estimation). 

Summarized results are illustrated in Figure \ref{fig.MAPEs4}. The RRM models have better predictions (i.e. less errors) for PD1, PD3, PD4, PD5, and PD8, while RUM models fit better for PD2, PD6, and PD7. Nonetheless, as can be seen, the average MAPEs of the RUM and RRM models are very close for all PDs (fluctuated lines). In fact, taking the averages of RUM's and RRM's average MAPEs (straight lines), values are about 13\% and 14\%. In other words, the RUM and RRM models' average accuracies are approximate 87\% and 86\%, respectively. As such, there is not much difference in the models' accuracies. This dataset can be estimated by either RUM or RRM models which provide comparable results. The similar performances of the two models are also confirmed in other studies, for instance \citet{hensher2013random,chorus2012random}.

\begin{figure} [ht]
	\centering
	\includegraphics[width=0.7\textwidth]{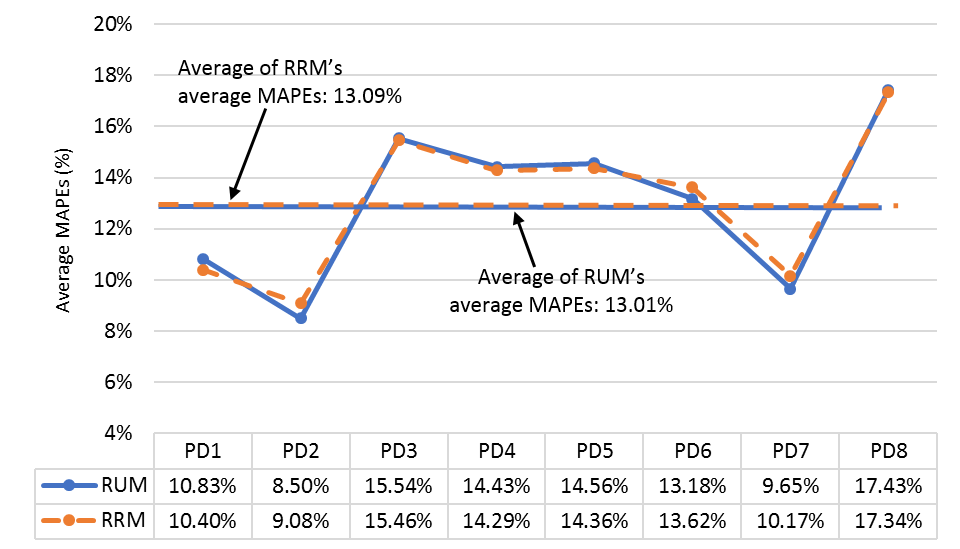}
	\caption{Average MAPEs of RUM and RRM models over PDs' testing samples}
	\label{fig.MAPEs4}	
\end{figure}\par

\section{Conclusions}
\noindent This study is the first investigation on senders' choice of couriers for the different PD types under the availability of both \<CS> and traditional carriers in a logistics market. Using RUM and RRM models to estimate the US survey data, we have revealed factors influencing senders' choice of courier service and the senders' WTP for different PDs. The shipping cost, delivery time, courier's reputation, tracking and tracing services, electronic delivery notification, and personalized delivery location and time window all may have a remarkable influence on senders' choices. Similarly, age, marital status, income, occupation, and number of family members may also significantly contribute to the senders' decisions. PDs which are found more likely to be sent via \<CS> systems include foods, beverages, and groceries. Apparel, books, music, videos, consumer electronics, and some other products are more likely to be sent via \<TLCs>, regardless of gender and age. Personal health and medicine deliveries are shared by both \<CS> and \<TLCs>. Interestingly, the WTP for shipping food, groceries, beverages, and medicine are found to be higher than those for other PDs. Moreover, a big adjustment in the shipping cost, which has the largest elasticity, needs to be made to change a sender's mind regarding couriers. Finally, to evaluate accuracies of the RUM and RRM models, we used data-science testing techniques. Those models found comparable performances even though the RUM models are better for predicting courier choice for sending groceries, books, music, videos, and consumer electronics.

Findings from this research provide a reliable source supporting for the pricing determination strategies of \<CS> firms. Moreover, the \<WTPs> are calculated for each \<PD> category and for each sub-service, such as reputation, tracking, e-notification, and personalization for delivery time and location, therefore, those values are useful for constructing a total service price. Also, \<CS> firms should change the level of service at least as its corresponding elasticity to expect changes in senders' decisions. 

In conclusion, insights from this study can supply new matrices to help both \<CS> and traditional logistics firms improve their services as well as to know which retailers to approach for marketing their services. The future research should investigate markets in which there is fierce competition between \<CS> and \<TLCs>, particularly where personal health and medicine PDs are more likely to be sent by \<CS> or \<TLCs>, and under what circumstances.

\newgeometry{top=2cm, bottom=2.5cm, left=1cm, right=1cm}

\appendix
\section{Appendix A}

\begin{table} [h]
\centering
\scriptsize
\caption{Variables and meanings}
\label{table.variablesoutputs}
\begin{tabular}{ll|ll}
\hline
\multicolumn{1}{c}{\textbf{Variable}} & \multicolumn{1}{c|}{\textbf{Meaning}}                                                                           & \multicolumn{1}{c}{\textbf{Variable}} & \multicolumn{1}{c}{\textbf{Meaning}}                                                              \\ \hline
COST                                  & \begin{tabular}[c]{@{}l@{}}Delivery cost (random parameter   - mean)\end{tabular}                             & AGE2                                  & 21-25 years old                                                                                   \\
NsCOST                                & \begin{tabular}[c]{@{}l@{}}Deliver cost (random parameter -   std)\end{tabular}                               & AGE3GEN                               & 26-30  years old male                                                                             \\
TIME                                  & Delivery time                                                                                                   & AGE5                                  & 36-40 years old                                                                                   \\
RANKING                               & Courier's reputation/ ranking                                                                                   & INCOME5                               & \begin{tabular}[c]{@{}l@{}}Personal annual income of \$100-\$150 million\end{tabular}          \\
TRACK                                 & Tracking and tracing ability                                                                                    & MARR                                  & Married                                                                                           \\
NOTI                                  & Electronic delivery notification                                                                                & MARROLD1                              & \begin{tabular}[c]{@{}l@{}}Married and living with people more than 64 years old\end{tabular} \\
TIWI                                  & \begin{tabular}[c]{@{}l@{}}Personalization for delivery time window\end{tabular}                            & LESS18                                & \begin{tabular}[c]{@{}l@{}}Living with people less than 18 years old\end{tabular}             \\
LOC2                                  & \begin{tabular}[c]{@{}l@{}}Personalization for location of   delivery \\(not home or pickup point)\end{tabular} & 18-64                                 & \begin{tabular}[c]{@{}l@{}}Living with people from 18 to 64 years old\end{tabular}            \\
TIP                                   & Willingness to tip                                                                                              & OCC1                                  & Full-time employees                                                                               \\
TIP1/A11\_Q381                        & \begin{tabular}[c]{@{}l@{}}Willingness to tip (for courier   1)\end{tabular}                                  & OCC                                   & Full- and part-time employees                                                                     \\
TIP2/B11\_Q382                        & \begin{tabular}[c]{@{}l@{}}Willingness to tip (for courier  2)\end{tabular}                                  & OCCGEN                                & Full-time male employees                                                                          \\
TIP3/C11\_Q383                        & \begin{tabular}[c]{@{}l@{}}Willingness to tip (for courier   3)\end{tabular}                                  & A\_A11/C\_A11                                & Constant - courier 1                                                                              \\
CONCERN1                              & \begin{tabular}[c]{@{}l@{}}Concerns on packages of being   damaged\end{tabular}                               & A\_B11/C\_B11                                & Constant - courier 2                                                                              \\
                                      &                                                                                                                 & A\_C11/C\_C11                                & Constant - courier 3                                                                              \\ \hline
\end{tabular}
\end{table}

\begin{figure} [h!]
	
	\centering
	\begin{subfigure}{.5\textwidth}
		\centering
		\includegraphics[height=0.65\textwidth]{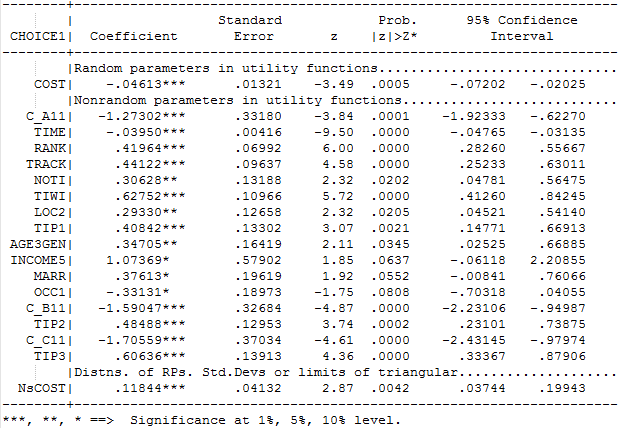}
		\captionof{figure}{PD1 RUM}
		\label{RUM_PD1}
	\end{subfigure}%
	\begin{subfigure}{.5\textwidth}
		\centering
		\includegraphics[height=0.65\textwidth]{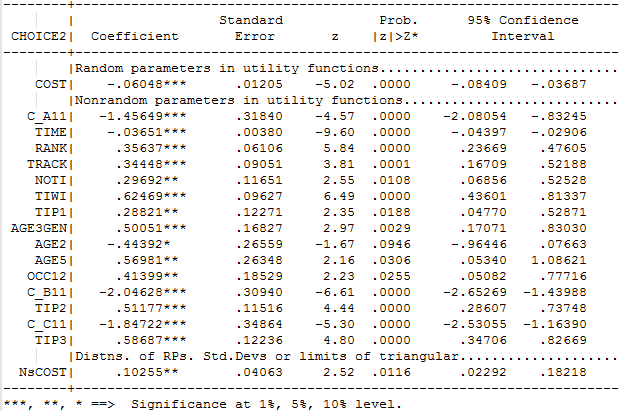}
		\captionof{figure}{PD2 RUM}
		\label{fig.RUM_PD2}
	\end{subfigure}
	
    \begin{subfigure}{.5\textwidth}
		\centering
		\includegraphics[height=0.65\textwidth]{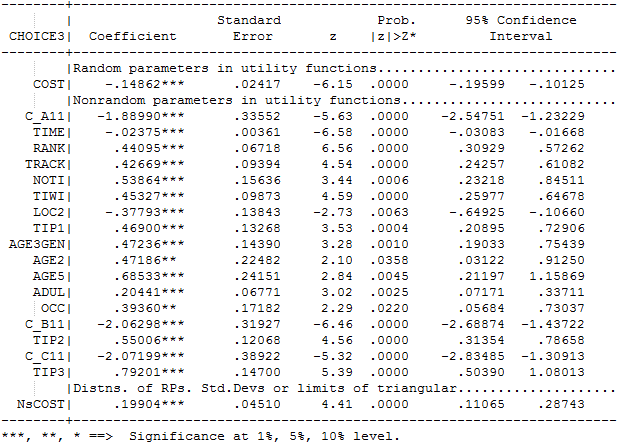}
		\captionof{figure}{PD3 RUM}
		\label{RUM_PD3}
	\end{subfigure}%
	\begin{subfigure}{.5\textwidth}
		\centering
		\includegraphics[height=0.65\textwidth]{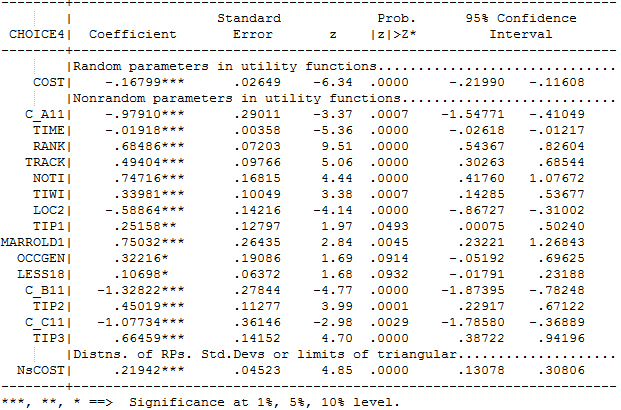}
		\captionof{figure}{PD4 RUM}
		\label{fig.RUM_PD4}
	\end{subfigure}   
    
    \addtocounter{figure}{-1}
    \captionof{figure}{Models' outputs}
	\label{fig.modelsoutputs}
\end{figure}

\begin{figure} \ContinuedFloat
	\begin{subfigure}{.5\textwidth}
		\centering
		\includegraphics[height=0.5\textwidth]{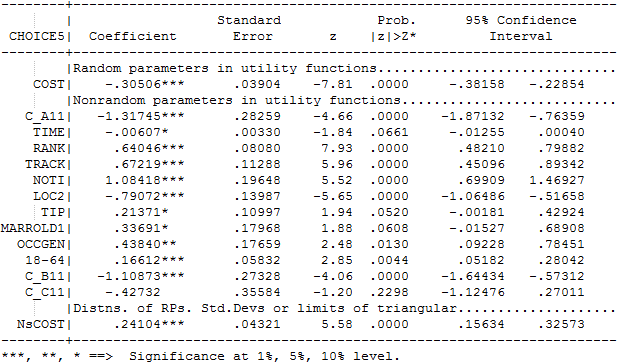}
		\captionof{figure}{PD5 RUM}
		\label{RUM_PD5}
	\end{subfigure}%
	\begin{subfigure}{.5\textwidth}
		\centering
		\includegraphics[height=0.5\textwidth]{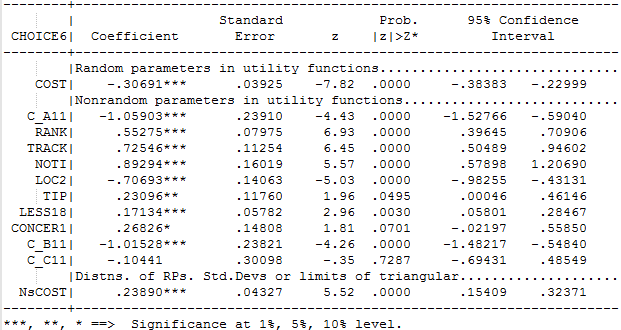}
		\captionof{figure}{PD6 RUM}
		\label{fig.RUM_PD6}
	\end{subfigure}
    \begin{subfigure}{.5\textwidth}
		\centering
		\includegraphics[height=0.5\textwidth]{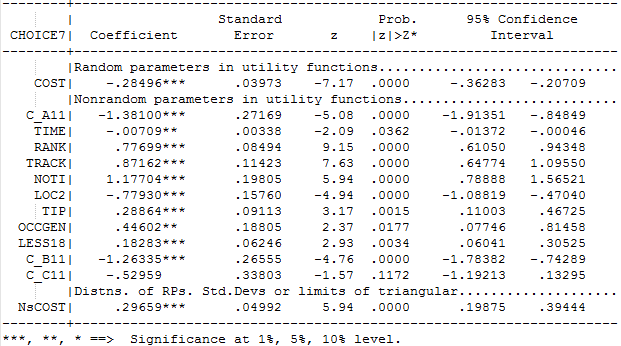}
		\captionof{figure}{PD7 RUM}
		\label{RUM_PD7}
	\end{subfigure}%
	\begin{subfigure}{.5\textwidth}
		\centering
		\includegraphics[height=0.5\textwidth]{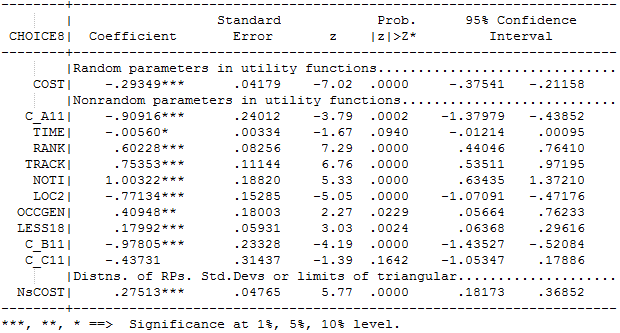}
		\captionof{figure}{PD8 RUM}
		\label{fig.RUM_PD8}
	\end{subfigure}
    
    \begin{subfigure}{.5\textwidth}
		\centering
		\includegraphics[height=0.5\textwidth]{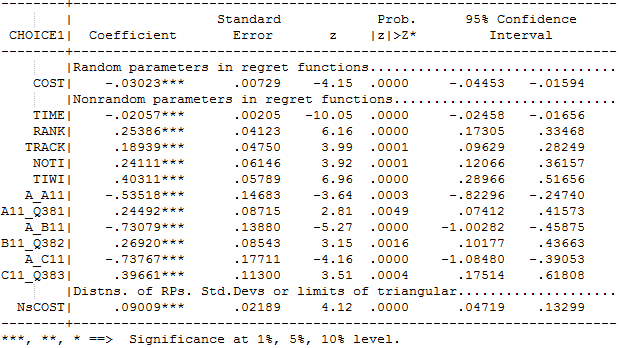}
		\captionof{figure}{PD1 RRM}
		\label{RRM_PD1}
	\end{subfigure}%
	\begin{subfigure}{.5\textwidth}
		\centering
		\includegraphics[height=0.5\textwidth]{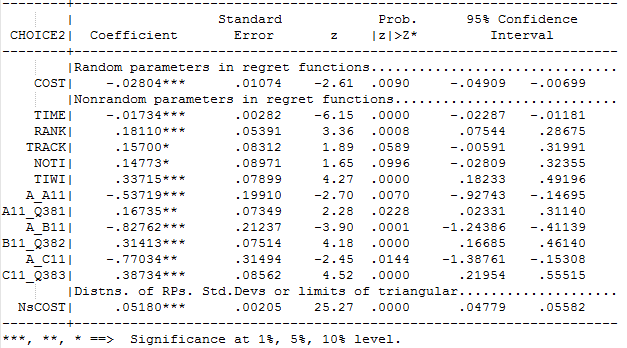}
		\captionof{figure}{PD2 RRM}
		\label{fig.RRM_PD2}
	\end{subfigure}
    
    \addtocounter{figure}{-1}
    \captionof{figure}{Models' outputs (Cont.)}
\end{figure}

\begin{figure} \ContinuedFloat
	\begin{subfigure}{.5\textwidth}
		\centering
        \includegraphics[height=0.5\textwidth]{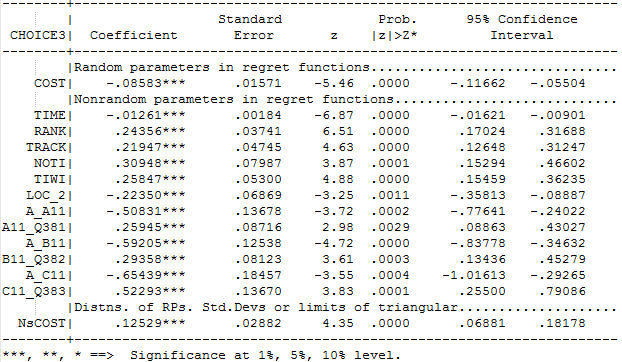}
		\captionof{figure}{PD3 RRM}
		\label{RRM_PD3}
	\end{subfigure}%
	\begin{subfigure}{.5\textwidth}
		\centering
		\includegraphics[height=0.5\textwidth]{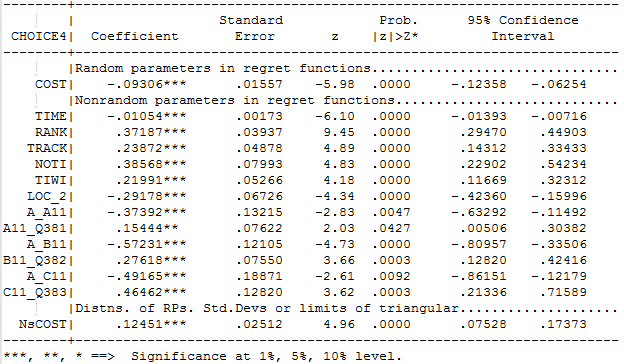}
		\captionof{figure}{PD4 RRM}
		\label{fig.RRM_PD4}
	\end{subfigure}
    
    \begin{subfigure}{.5\textwidth}
		\centering
		\includegraphics[height=0.45\textwidth]{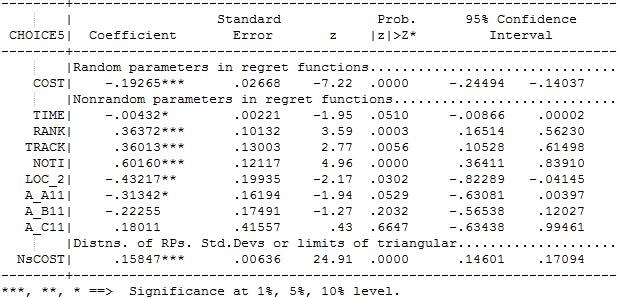}
		\captionof{figure}{PD5 RRM}
		\label{RRM_PD5}
	\end{subfigure}%
	\begin{subfigure}{.5\textwidth}
		\centering
		\includegraphics[height=0.45\textwidth]{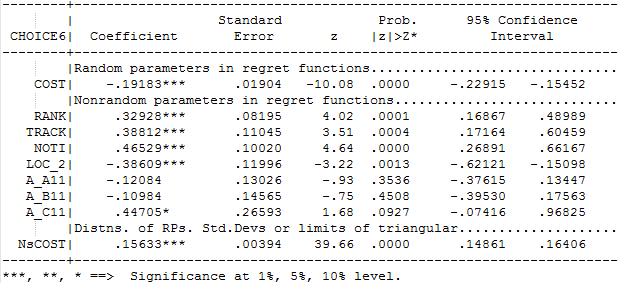}
		\captionof{figure}{PD6 RRM}
		\label{fig.RRM_PD6}
	\end{subfigure}
    \begin{subfigure}{.5\textwidth}
		\centering
		\includegraphics[height=0.45\textwidth]{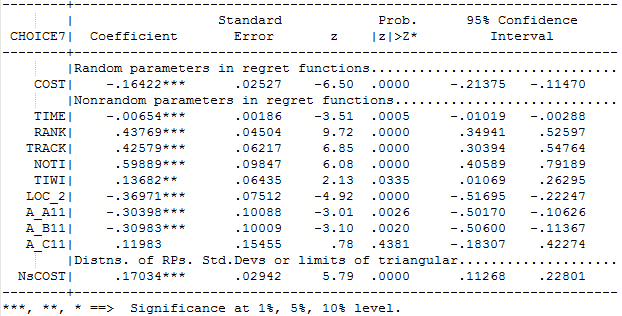}
		\captionof{figure}{PD7 RRM}
		\label{RRM_PD7}
	\end{subfigure}%
	\begin{subfigure}{.5\textwidth}
		\centering
		\includegraphics[height=0.45\textwidth]{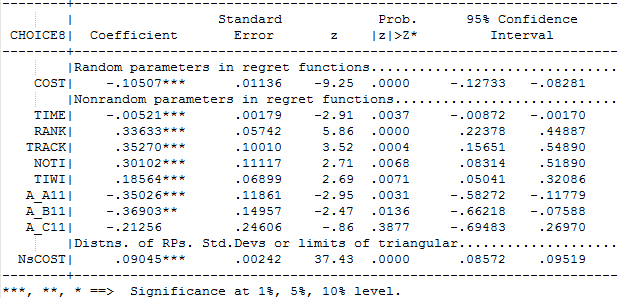}
		\captionof{figure}{PD8 RRM}
		\label{fig.RRM_PD8}
	\end{subfigure}
    
    \addtocounter{figure}{-1}
    \captionof{figure}{Models' outputs (Cont.)}
\end{figure} 

\restoregeometry

\pagebreak
\section{Appendix B}
\begin{figure} [h!]	
	\centering
	\begin{subfigure}{.5\textwidth}
		\centering
		\includegraphics[height=0.5\textwidth]{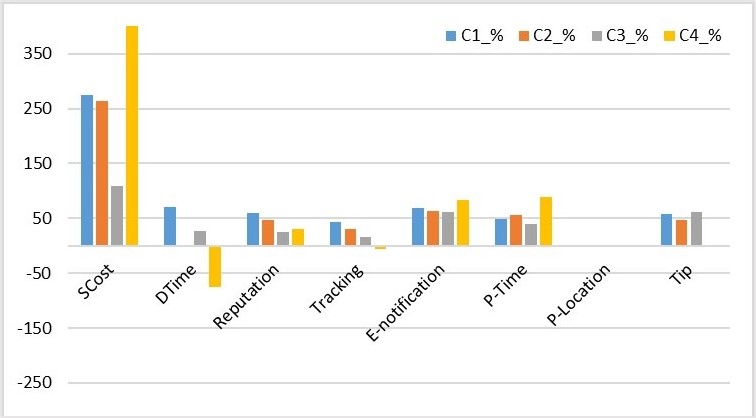}
		\captionof{figure}{PD1's elasticity differences (\%)}
		\label{elas_PD1}
	\end{subfigure}%
	\begin{subfigure}{.5\textwidth}
		\centering
		\includegraphics[height=0.5\textwidth]{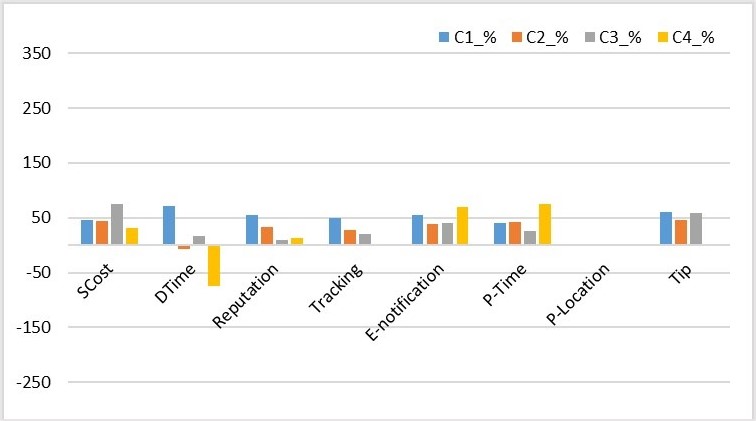}
		\captionof{figure}{PD2's elasticity differences (\%)}
		\label{fig.elas_PD2}
	\end{subfigure}
	
	\begin{subfigure}{.5\linewidth}
		\centering
		\includegraphics[height=0.5\linewidth]{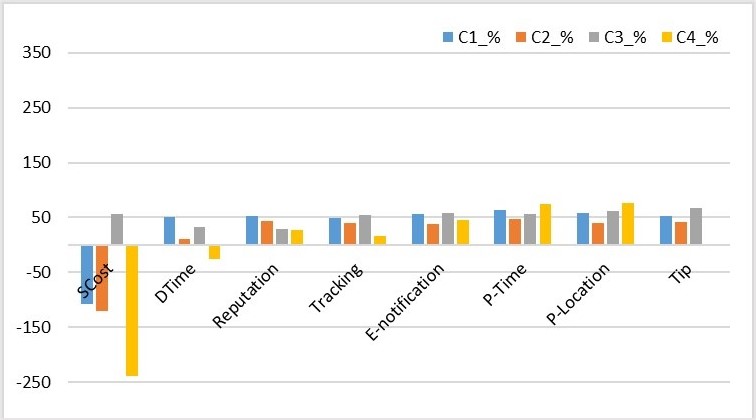}
		\captionof{figure}{PD3's elasticity differences (\%)}
		\label{fig.elas_PD3}
	\end{subfigure}%
	\begin{subfigure}{.5\linewidth}
		\centering
		\includegraphics[height=0.5\linewidth]{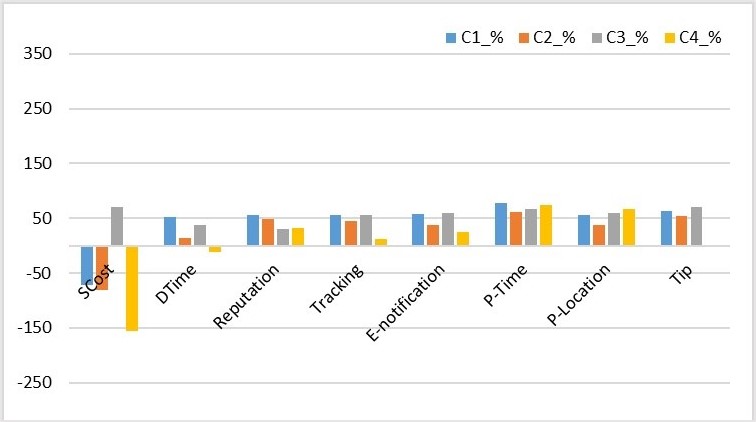}
		\captionof{figure}{PD4's elasticity differences (\%)}
		\label{fig.elas_PD4}
		
	\end{subfigure}
	\begin{subfigure}{.5\linewidth}
		\centering
		\includegraphics[height=0.5\linewidth]{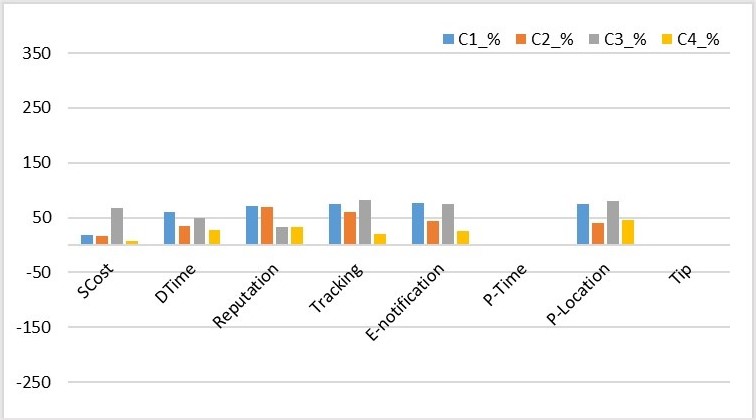}
		\captionof{figure}{PD5's elasticity differences (\%)}
		\label{fig.elas_PD5}
	\end{subfigure}%
	\begin{subfigure}{.5\linewidth}
		\centering
		\includegraphics[height=0.5\linewidth]{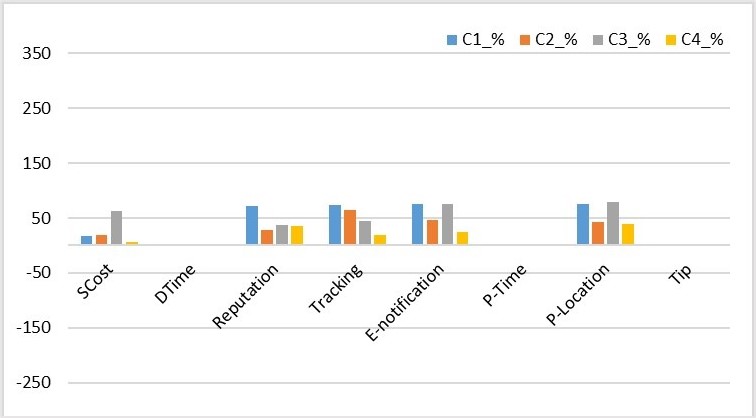}
		\captionof{figure}{PD6's elasticity differences (\%)}
		\label{fig.elas_PD6}
	\end{subfigure}
	
	\begin{subfigure}{.5\linewidth}
		\centering
		\includegraphics[height=0.5\linewidth]{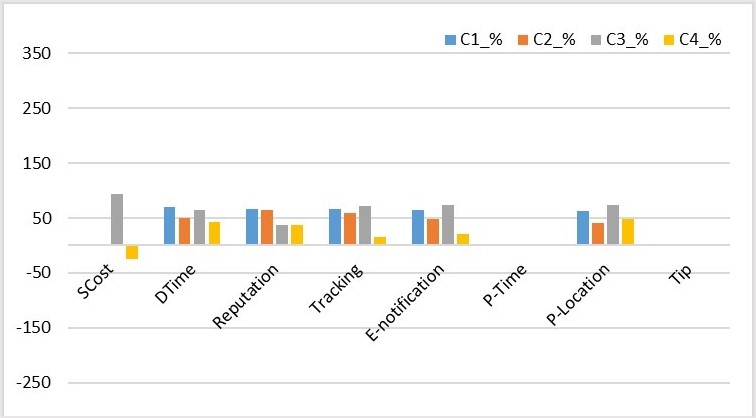}
		\captionof{figure}{PD7's elasticity differences (\%)}
		\label{fig.elas_PD7}
	\end{subfigure}%
	\begin{subfigure}{.5\linewidth}
		\centering
		\includegraphics[height=0.5\linewidth]{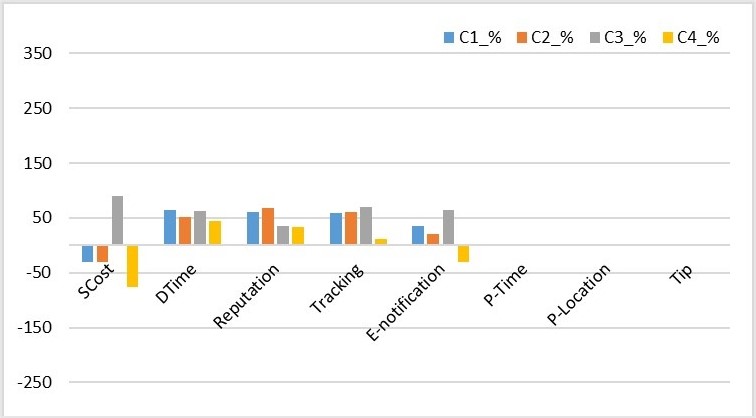}
		\captionof{figure}{PD8's elasticity differences (\%)}
		\label{fig.elas_PD8}
	\end{subfigure}
    
    \addtocounter{figure}{-1}
	\captionof{figure}{Differences of RUM and RRM models' elasticities on alternatives' attributes}
	\label{fig.elas4}
\end{figure}

\noindent \textbf{AUTHOR CONTRIBUTION STATEMENT} \\
The authors confirm contribution to the paper as follows: study conception and design: Le, Ukkusuri; data collection: Le, Ukkusuri; analysis and interpretation of results: Le; draft manuscript preparation: Le, Ukkusuri. All authors reviewed the results and approved the final version of the manuscript.

\bibliographystyle{trb}
\bibliography{citationrp}

\begin{thebibliography}{39}
\providecommand{\natexlab}[1]{#1}

\bibitem[{Dablanc(2016)}]{Dablanc2016}
Dablanc, L., Digital market places for urban freight: is digital city logistics
  a disruption to the urban freight routine? In \emph{Presentation in The VREF
  Conference on Urban Freight 2016: Plan for the future sharing urban space
  17–19 October 2016, Gothenburg.}, 2016.

\bibitem[{Li et~al.(2015)Li, Moreno, and Zhang}]{li2015agent}
Li, J., A.~Moreno, and D.~J. Zhang, Agent behavior in the sharing economy:
  Evidence from Airbnb. \emph{Ross School of Business Working Paper Series}, ,
  No. 1298, 2015.

\bibitem[{Ke(2017)}]{ke2017service}
Ke, Q., Service Providers of the Sharing Economy: Who Joins and Who Benefits?
  \emph{arXiv preprint arXiv:1709.07580}, 2017.

\bibitem[{Zervas et~al.(2017)Zervas, Proserpio, and Byers}]{zervas2017rise}
Zervas, G., D.~Proserpio, and J.~W. Byers, The rise of the sharing economy:
  Estimating the impact of Airbnb on the hotel industry. \emph{Journal of
  Marketing Research}, Vol.~54, No.~5, 2017, pp. 687--705.

\bibitem[{Statista(2018)}]{Statista2018}
Statista, Company value and equity funding of Airbnb from 2014 to 2017 (in
  billion U.S. dollars). \emph{Statista}, 2018.

\bibitem[{Isaac(2017)}]{Isaac2017}
Isaac, M., One Surprise Standout for Uber: Food Delivery. \emph{The New York
  Times}, 2017.

\bibitem[{Dervojeda(2013)}]{dervojeda2013sharing}
Dervojeda, K., \emph{The sharing economy: accessibility based business models
  for peer-to-peer markets}. European Commission, 2013.

\bibitem[{Schor(2016)}]{schor2016debating}
Schor, J., DEBATING THE SHARING ECONOMY. \emph{Journal of Self-Governance \&
  Management Economics}, Vol.~4, No.~3, 2016.

\bibitem[{Frenken and Schor(2017)}]{frenken2017putting}
Frenken, K. and J.~Schor, Putting the sharing economy into perspective.
  \emph{Environmental Innovation and Societal Transitions}, Vol.~23, 2017, pp.
  3--10.

\bibitem[{Train and Wilson(2008)}]{train2008estimation}
Train, K. and W.~W. Wilson, Estimation on stated-preference experiments
  constructed from revealed-preference choices. \emph{Transportation Research
  Part B: Methodological}, Vol.~42, No.~3, 2008, pp. 191--203.

\bibitem[{Anderson et~al.(2009)Anderson, Opaluch, and
  Grigalunas}]{anderson2009demand}
Anderson, C.~M., J.~J. Opaluch, and T.~A. Grigalunas, The demand for import
  services at US container ports. \emph{Maritime Economics \& Logistics},
  Vol.~11, No.~2, 2009, pp. 156--185.

\bibitem[{Danielis and Marcucci(2007)}]{danielis2007attribute}
Danielis, R. and E.~Marcucci, Attribute cut-offs in freight service selection.
  \emph{Transportation Research Part E: Logistics and Transportation Review},
  Vol.~43, No.~5, 2007, pp. 506--515.

\bibitem[{Zamparini et~al.(2011)Zamparini, Layaa, and
  Dullaert}]{zamparini2011monetary}
Zamparini, L., J.~Layaa, and W.~Dullaert, Monetary values of freight transport
  quality attributes: A sample of Tanzanian firms. \emph{Journal of Transport
  Geography}, Vol.~19, No.~6, 2011, pp. 1222--1234.

\bibitem[{Cavalcante and Roorda(2013)}]{cavalcante2013shipper}
Cavalcante, R.~A. and M.~J. Roorda, Shipper/carrier interactions data
  collection: Web-based respondent customized stated preference (WRCSP) survey.
  In \emph{Transport Survey Methods: Best Practice for Decision Making},
  Emerald Group Publishing Limited, 2013, pp. 257--278.

\bibitem[{Joerss et~al.(2016)Joerss, Schröder, Neuhaus, Klink, and
  Mann}]{Joerss2016}
Joerss, M., J.~Schröder, F.~Neuhaus, C.~Klink, and F.~Mann, \emph{Parcel
  delivery: The future of last mile}. McKinsey\&Company, 2016.

\bibitem[{Tussyadiah(2016)}]{tussyadiah2016factors}
Tussyadiah, I.~P., Factors of satisfaction and intention to use peer-to-peer
  accommodation. \emph{International Journal of Hospitality Management},
  Vol.~55, 2016, pp. 70--80.

\bibitem[{AirBnBcitizen(2017)}]{AirBnBcitizen2017}
AirBnBcitizen, Airbnb’s 2016 highlights and 2017 trends we’re watching.
  \emph{AirBnBcitizen}, 2017.

\bibitem[{Guttentag et~al.(2017)Guttentag, Smith, Potwarka, and
  Havitz}]{guttentag2017tourists}
Guttentag, D., S.~Smith, L.~Potwarka, and M.~Havitz, Why tourists choose
  Airbnb: a motivation-based segmentation study. \emph{Journal of Travel
  Research}, 2017, p. 0047287517696980.

\bibitem[{Oliphant(2008)}]{oliphant2008native}
Oliphant, M., \emph{The native slugs of northern Virginia: A profile of
  slugging in the Washington DC region}. Master's thesis, Master of Sciences in
  Urban and Regional Planning (Blacksburg, VA: Virginia Polytechnic Institute),
  2008.

\bibitem[{Rayle et~al.(2016)Rayle, Dai, Chan, Cervero, and
  Shaheen}]{rayle2016just}
Rayle, L., D.~Dai, N.~Chan, R.~Cervero, and S.~Shaheen, Just a better taxi? A
  survey-based comparison of taxis, transit, and ridesourcing services in San
  Francisco. \emph{Transport Policy}, Vol.~45, 2016, pp. 168--178.

\bibitem[{Shaheen et~al.(2016)Shaheen, Chan, and Gaynor}]{shaheen2016casual}
Shaheen, S.~A., N.~D. Chan, and T.~Gaynor, Casual carpooling in the San
  Francisco Bay Area: Understanding user characteristics, behaviors, and
  motivations. \emph{Transport Policy}, Vol.~51, 2016, pp. 165--173.

\bibitem[{Dias et~al.(2017)Dias, Lavieri, Garikapati, Astroza, Pendyala, and
  Bhat}]{dias2017behavioral}
Dias, F.~F., P.~S. Lavieri, V.~M. Garikapati, S.~Astroza, R.~M. Pendyala, and
  C.~R. Bhat, A behavioral choice model of the use of car-sharing and
  ride-sourcing services. \emph{Transportation}, Vol.~44, No.~6, 2017, pp.
  1307--1323.

\bibitem[{Smith(2016)}]{Smith2016}
Smith, A., Shared, collaborative and on demand: the new digital economy.
  \emph{{Pew Research Center}}, 2016.

\bibitem[{Carrese et~al.(2017)Carrese, Giacchetti, Patella, and
  Petrelli}]{carrese2017real}
Carrese, S., T.~Giacchetti, S.~Patella, and M.~Petrelli, Real time ridesharing:
  Understanding user behavior and policies impact: Carpooling service case
  study in Lazio Region, Italy. In \emph{Models and Technologies for
  Intelligent Transportation Systems (MT-ITS), 2017 5th IEEE International
  Conference on}, IEEE, 2017, pp. 721--726.

\bibitem[{Bennett(2015)}]{Bennett2015}
Bennett, J., Uber Drives Data. \emph{5isolution}, 2015.

\bibitem[{Kooti et~al.(2017)Kooti, Grbovic, Aiello, Djuric, Radosavljevic, and
  Lerman}]{kooti2017analyzing}
Kooti, F., M.~Grbovic, L.~M. Aiello, N.~Djuric, V.~Radosavljevic, and
  K.~Lerman, Analyzing Uber's Ride-sharing Economy. In \emph{Proceedings of the
  26th International Conference on World Wide Web Companion}, International
  World Wide Web Conferences Steering Committee, 2017, pp. 574--582.

\bibitem[{Miller et~al.(2017)Miller, Nie, and
  Stathopoulos}]{miller2017crowdsourced}
Miller, J., Y.~Nie, and A.~Stathopoulos, Crowdsourced Urban Package Delivery:
  Modeling Traveler Willingness to Work as Crowdshippers. \emph{Transportation
  Research Record: Journal of the Transportation Research Board}, , No. 2610,
  2017, pp. 67--75.

\bibitem[{Punel and Stathopoulos(2017)}]{punel2017modeling}
Punel, A. and A.~Stathopoulos, Modeling the acceptability of crowdsourced goods
  deliveries: Role of context and experience effects. \emph{Transportation
  Research Part E: Logistics and Transportation Review}, Vol. 105, 2017, pp.
  18--38.

\bibitem[{Briffaz and Darvey(2016)}]{briffaz2016crowd}
Briffaz, M. and C.~Darvey, \emph{Crowd-shipping in Geneva Exploratory and
  descriptive study of Crowd-shipping}. Master's thesis, University of
  Gothenburg, 2016.

\bibitem[{Le and Ukkusuri(2018)}]{Le2017}
Le, T.~V. and S.~V. Ukkusuri, Crowd-shipping services for last mile delivery:
  analysis from survey data in two countries. In \emph{Transportation Research
  Board 97th Annual Meeting, Washington, D.C., Jan. 7–11, paper no.
  18-03779}, 2018.

\bibitem[{Ballare and Lin(2018)}]{ballare2018preliminary}
Ballare, S. and J.~Lin, Preliminary Investigation of a Crowdsourced Package
  Delivery System: A Case Study. \emph{City Logistics 3: Towards Sustainable
  and Liveable Cities}, 2018, pp. 109--128.

\bibitem[{Punel et~al.(2018)Punel, Ermagun, and Stathopoulos}]{punel2018}
Punel, A., A.~Ermagun, and A.~Stathopoulos, Studying Determinants of
  Crowd-shipping Use. \emph{Travel Behavior and Society}, Vol.~12, 2018, pp. 30
  -- 40.

\bibitem[{{ATKearney}(2014)}]{ATKearney2014a}
{ATKearney}, \emph{Connected consumers are not created equal: A global
  perspective}, 2014, accessed 2017-06-01.

\bibitem[{Chorus et~al.(2008)Chorus, Arentze, and
  Timmermans}]{chorus2008random}
Chorus, C.~G., T.~A. Arentze, and H.~J. Timmermans, A random
  regret-minimization model of travel choice. \emph{Transportation Research
  Part B: Methodological}, Vol.~42, No.~1, 2008, pp. 1--18.

\bibitem[{Small(2012)}]{small2012valuation}
Small, K.~A., Valuation of travel time. \emph{Economics of transportation},
  Vol.~1, No.~1, 2012, pp. 2--14.

\bibitem[{Hensher et~al.(2013)Hensher, Greene, and Chorus}]{hensher2013random}
Hensher, D.~A., W.~H. Greene, and C.~G. Chorus, Random regret minimization or
  random utility maximization: an exploratory analysis in the context of
  automobile fuel choice. \emph{Journal of Advanced Transportation}, Vol.~47,
  No.~7, 2013, pp. 667--678.

\bibitem[{Chorus(2012)}]{chorus2012random}
Chorus, C., Random regret minimization: an overview of model properties and
  empirical evidence. \emph{Transport Reviews}, Vol.~32, No.~1, 2012, pp.
  75--92.

\bibitem[{Roug{\`e}s and Montreuil(2014)}]{rouges2014crowdsourcing}
Roug{\`e}s, J.-F. and B.~Montreuil, Crowdsourcing delivery: New interconnected
  business models to reinvent delivery. In \emph{1st international physical
  internet conference}, 2014, pp. 1--19.

\bibitem[{Carbone et~al.(2015)Carbone, Rouquet, and
  Roussat}]{carbone2015carried}
Carbone, V., A.~Rouquet, and C.~Roussat, Carried away by the crowd”: what
  types of logistics characterise collaborative consumption. In \emph{1st
  International Workshop on Sharing Economy}, 2015, pp. 1--21.

\end{thebibliography}

\end{document}